\RequirePackage{fix-cm}

\documentclass[twocolumn]{svjour3}          
\smartqed  
\usepackage{graphicx}

\usepackage{enumitem}

\usepackage{graphicx}
\usepackage{balance}  

\usepackage{multirow}
\usepackage{listings}
\usepackage{graphicx}
\usepackage{url}
\usepackage{algorithm}

\usepackage{float}
\newfloat{algorithm}{t}{lop}

\usepackage{algpseudocode}
\usepackage{xcolor}
\usepackage{balance}

\usepackage{color, colortbl}
\usepackage{subfigure}
\usepackage[T1]{fontenc}
\usepackage{tikz}
\newcommand*\circled[1]{\tikz[baseline=(char.base)]{
                \node[shape=circle,draw,inner sep=0.5pt] (char) {#1};}}

              \usepackage{caption}

\newcommand{\cred}{}
\newcommand{\cb}{}
\newcommand{\cg}{}

\begin{document}
\title{\cred{TADOC: Text Analytics Directly on Compression}}



\author{
}

\author{
  Feng Zhang{\small$^{1}$} \and
Jidong Zhai{\small$^{2}$} \and
Xipeng Shen{\small$^{3}$} \and
  Dalin Wang{\small$^{1}$} \and
\\
  Zheng Chen{\small$^{1}$} \and
Onur Mutlu{\small$^{4}$} \and
Wenguang Chen{\small$^{2}$} \and
Xiaoyong Du{\small$^{1}$}
}


\authorrunning{Feng Zhang, Jidong Zhai, Xipeng Shen, et al.}

\institute{Feng Zhang \at
              \email{fengzhang@ruc.edu.cn}           
           \and
           Jidong Zhai \at
              \email{zhaijidong@tsinghua.edu.cn}           
           \and
           Xipeng Shen \at
              \email{xshen5@ncsu.edu}           
           \and
           Dalin Wang \at
              \email{sxwangdalin@ruc.edu.cn}           
           \and
           Zheng Chen  \at
              \email{2016202201@ruc.edu.cn}           
           \and
           Onur Mutlu \at
              \email{onur.mutlu@inf.ethz.ch}           
           \and
           Wenguang Chen \at
              \email{cwg@tsinghua.edu.cn}           
           \and
           Xiaoyong Du \at
              \email{duyong@ruc.edu.cn}           
              \and\at
              {\small$~^{1}$}DEKE Lab, School of Information, Renmin University of China, China 
              \and\at
              {\small$~^{2}$}Department of Computer Science and Technology, Tsinghua University, China  
              \and\at
{\small$~^{3}$}Computer Science Department, North Carolina State University, USA  
              \and\at
{\small $~^{4}$}Department of Computer Science, ETH Z{\"u}rich, Switzerland 
}

\date{Received: date / Accepted: date}

\maketitle

\def\parallelAVG{2.4X}

\def\singleAVG{1.6X}   
\def\singleAVGa{2.0X}
\def\singleAVGb{1.9X}
\def\singleAVGplusa{3.2X}

\def\singleHigha{2.8X}
\def\singleHighb{3.0X}
\def\singleAVGplusb{2.3X}

\def\gzipSlowDown{19.8\%}

\def\init{2.4X}
\def\compute{1.6X}

\def\totalAVG{2.0X}

\def\multiAVGplus{1.1X}
\def\multiAVG{1.4X}
\def\multiHigh{2.7X}

\def\distriAVG{2.2X}  
\def\distriAVGplus{1.9X}
\def\distriHigh{2.1X}
\def\gzipCompressRatio{14.16\%}
\def\naiveCompressRatio{39.74\%}
\def\advanceCompressRatio{11.86\%}

\def\compressRatioAVG{11.86\%}
\def\reduceRatioAVG{90.8\%}
\def\memSavingAVG{87.9\%}

\def\wikiCompressSaving{64.1\%}

\def\wordCtvmem1{35.9\%}
\def\wordCountv1timeper{56.8\%}
\def\wordCtvsecondf{1.8X}
\def\wordCtv2per{1\%}
\def\wordCthirdSpeedup{2.8X}


\def\basicZwiftNSFRAA{2.3}
\def\basicZwiftWiki{2.8}
\def\ZwiftNSFRAA{6.5}
\def\ZwiftWiki{11.9}
\def\ZwiftAverage{11.8}
\def\ZavgSpeedupA{2.33}
\def\ZavgSpeedupB{2.22}
\def\ZavgSpeedupC{2.71}

\def\ZavgSpeedupAadd{2.08}
\def\ZavgSpeedupBadd{2.12}
\def\ZavgSpeedupCadd{2.59}

\begin{abstract}


  \cred{
This article provides a comprehensive description of Text Analytics Directly on Compression (TADOC), which enables direct document analytics on compressed textual data. The article explains the concept of TADOC and the challenges to its effective realizations.
  Additionally, a series of guidelines and technical solutions that effectively address those challenges, including the adoption of a hierarchical compression method and a set of novel algorithms and data structure designs, are presented. 
  Experiments on six data analytics tasks of various complexities show that TADOC can save  \reduceRatioAVG{} storage space and \memSavingAVG{} memory usage, while halving data processing times. 
}

  
\end{abstract}


\vspace{-0.40in}

\section{Introduction}
\label{sec:intro}
\vspace{-0.10in}

\emph{Document analytics} refers to data analytics tasks that derive
statistics, patterns, insights or knowledge from textual
documents (e.g., system log files, emails, corpus). It is important for 
many applications, from web search to system diagnosis,
security, and so on.
Document analytics applications are time-consuming, especially as the
data they process keep growing rapidly. At the same time, they often
need a large amount of space, both in storage and memory.

A common approach to mitigating the space concern is data compression. Although it often reduces the storage usage by several factors,
compression does \emph{not} alleviate, but actually worsens, the time
concern. In current document analytics frameworks, compressed documents
have to be decompressed before being processed. The decompression
step lengthens the end-to-end processing time. 

This work investigates the feasibility of efficient data analytics on
compressed data \emph{without} decompressing it. Its motivation is
two-fold. First, it could avoid the decompression time. Second, more
importantly, it could save some processing. Space savings by
compression fundamentally stems from repetitions in the data. If the
analytics algorithms could leverage the repetitions that the
compression algorithm already uncovers, it could avoid 
unnecessary repeated processing, and hence shorten the processing
time significantly. Compression takes time. But many datasets (e.g.,
government document archives, electronic book collections, historical
Wikipedia datasets \cite{wikipedia}) are used for various analytics
tasks by many users repeatedly. For them, the compression time is well
justified by the repeated usage of the compression results.

\cred{
This article presents Text Analytics Directly on Compression (TADOC), which  enables direct document analytics on compressed textual data. 
}
We base
TADOC on a specific compression algorithm named {\em
  Sequitur}~\cite{nevill1997identifying} for the hierarchical
structure of its compression results (Section~\ref{sec:premise}).

We introduce the concept of {\em compression-based direct
  processing}, and analyze its challenges
(Section~\ref{sec:challenges}).
Through studies on a set of core algorithms used in document
analytics, we discover a set of solutions and insights on tackling
those challenges. These insights range from algorithm designs to data
structure selections, scalable implementations, and adaptations to
various problems and datasets. We draw on several common document analytics
problems to explain our
insights, and provide the first set of essential guidelines and techniques for
effective compression-based document analytics
(Section~\ref{sec:solutions}).

Our work yields an immediately-usable artifact, the \\
\texttt{CompressDirect} library, which
offers a set of modules to ease the application of our guidelines.
Our library provides implementations of six
algorithms frequently used in document analytics, in sequential,
parallel, and distributed versions, which can be directly plugged into
existing applications to generate immediate benefits.

\cred{
We further discuss how TADOC and its associated CompressDirect library can be effectively applied to real-world data analytics applications. We demonstrate the process on four applications,
{\em word co-occurrence}~\cite{matsuo2004keyword,pennington2014glove},
\ 
{\em term frequency-inverse document frequency}~\cite{joachims1996probabilistic},
{\em word2vec}~\cite{rong2014word2vec,googleWord2Vec}, 
and {\em latent Dirichlet allocation} (LDA) \cite{blei2003latent}.
Our evaluation validates the efficacy of our proposed techniques in 
saving both space and time on six analytics kernels.  
For six common analytics kernels, compared to their default implementation on uncompressed datasets, TADOC reduces
storage usage by
\reduceRatioAVG{} and memory usage by \memSavingAVG{}, and at the same
time, speeds up the analytics by \singleAVG{} for sequential runs,
and by \distriAVG{} for Spark-based
distributed runs.
On four real-world applications,
TADOC reduces  storage usage by 92.4\% and  memory usage by 26.1\%,
and yields 1.2X speedup over the original applications, on average.
}

A prior work, Succinct~\cite{agarwal2015succinct}, offers a way to enable efficient queries on compressed data. This work complements it by making complex document analytics on compressed data efficiently.
Data deduplication~\cite{monge1996field} saves storage
space, but does \emph{not} save repeated processing of the data.
\cg{
Our preliminary work has been presented in~\cite{zhang2018efficient}.
}


Overall, this work makes the following contributions:

\begin{itemize}[noitemsep,nolistsep]

  \item \cred{It presents an effective method for enabling high performance
    complex document analytics directly on compressed data, and
    realizes the method on the {\em Sequitur} compression algorithm.
    }

  \item It unveils the challenges of performing {\em compression-based
    document analytics} and offers a set of solutions, insights, and guidelines.

  \item It validates the efficacy of the proposed techniques, demonstrates
    their significant benefits in both space and time savings, and offers 
    a library for supporting common
    operations in document analytics. 

  \item 
    \cred{
      It demonstrates that 
     TADOC can be effectively applied to real-world document analytics applications,
     bringing 92.4\% storage space savings and 1.2X performance improvement. 
   }
\end{itemize}

\vspace{-0.25in}
\section{Premises and Background}
\label{sec:premise}
\vspace{-0.12in}

\cred{
  Operating directly on grammar-compressed data is an active research area in recent years~\cite{navarro2016compact,sadakane2007succinct}, which will be discussed in Section~\ref{sec:related}.
In this section, we present the premises and background of Sequitur compression algorithm and typical document analytics.
}

\vspace{-0.10in}
\vspace{-0.15in}

\subsection{Sequitur Algorithm}
\label{subsec:sequitur}
\vspace{-0.15in}

There are many compression algorithms for documents, such as
LZ77~\cite{ziv1977universal}, suffix array~\cite{navarro2016compact},
and their variants.  Our study focuses on
Sequitur~\cite{nevill1997identifying}
since its compression results are a natural fit for direct processing.

Sequitur is a recursive algorithm that infers a hierarchical
structure from a sequence of discrete symbols. For a given sequence of
symbols, it derives a context-free grammar (CFG), with each rule in
the CFG reducing a repeatedly appearing string into a single rule
ID. By replacing the original string with the rule ID in the CFG,
Sequitur makes it output CFG more compact than the original dataset.

\begin{figure}
  \center
\includegraphics[width=1\linewidth]{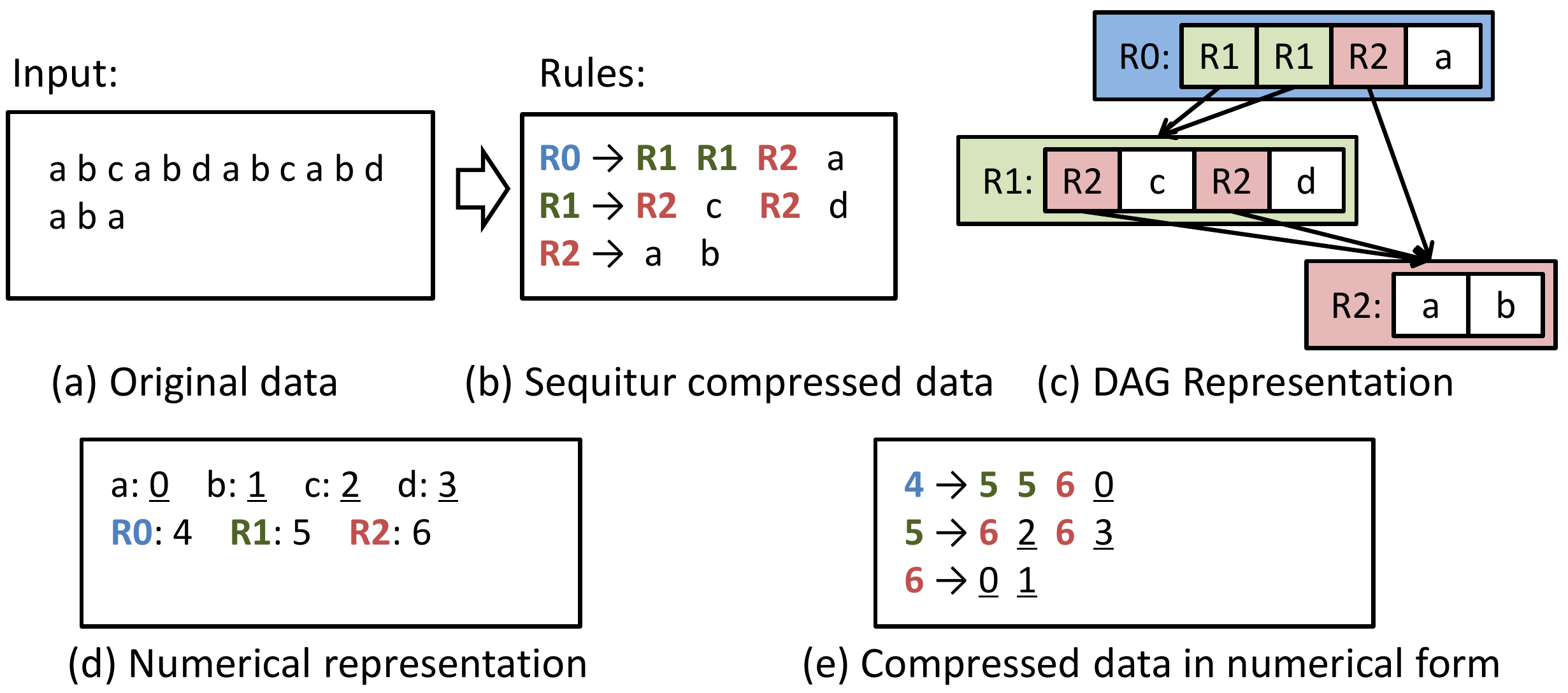}
\vspace{-0.24in}
  \caption{A compression example with Sequitur.}
  \label{fig:SequiturExample}
\vspace{0.01in}
\vspace{0.01in}
\vspace{-0.1in}
\vspace{-2.0mm}
\vspace{-2.0mm}
\end{figure}

Figure~\ref{fig:SequiturExample} illustrates Sequitur compression results.
Figure~\ref{fig:SequiturExample} (a)
shows the original input, and Figure~\ref{fig:SequiturExample} (b)
shows the output of Sequitur in a CFG form. The CFG uncovers both the
repetitions in the input string and the hierarchical structure.
It uses R0 to represent the entire string, which
consists of substrings represented by R1 and R2. The two instances of
R1 in R0 reflect the repetition of ``a b c a b d'' in the input
string, while the two instances of R2 in R1 reflect the
repetition of ``a b'' in the substring of R1.  The output of Sequitur
is often visualized with a directed acyclic graph (DAG), as
Figure~\ref{fig:SequiturExample} (c) shows. 
The edges indicate the
hierarchical relations among the rules. 

Dictionary encoding is often used to represent each word with a unique non-negative
integer.
A dictionary stores the mapping between integers and words. 
We represent each rule ID with a unique integer greater than $N$,
where $N$ is the total number of unique words contained in the
dataset. Figure~\ref{fig:SequiturExample} (d) gives the numerical
representations of the words and rules in
Figure~\ref{fig:SequiturExample} (a,b), while
Figure~\ref{fig:SequiturExample} (e) shows the CFG in numerical
form.




Sequitur provides compression ratios similar to those of other popular algorithms (e.g., Gzip)~\cite{nevill1996inferring}.
  Its compression process is relatively slow, but our technique is designed for datasets that are \emph{repeatedly} used by many users. For them, compression time
  is \emph{not} a main concern as the compression results can be used many times by different users for various tasks. Such datasets are common, ranging from book collections to historical Wikipedia pages~\cite{wikipedia}, government document archives, archived collections (e.g., of a law firm), historical news collections, and so on. 

Sequitur has several properties that make it appealing for our
use. First, the CFG structure in its results makes it easy to find
repetitions in input strings. Second, its output consists of the
direct (sequences of) input symbols rather than other indirect coding
  of the input (e.g., \emph{distance} used in LZ77~\cite{ziv1977universal}
and suffix array~\cite{navarro2016compact}). These properties make
Sequitur a good fit for materializing the idea of
compression-based direct processing. 


\vspace{-0.02in}
\vspace{-0.20in}
\subsection{Typical Document Analytics}
\label{sec:example}
\vspace{-0.01in}
\vspace{-0.10in}
Before presenting the proposed technique, we first describe three
commonly-performed document analytics tasks. They each feature different
challenges that are typical to many document analytics, offering
examples we use in later sections for discussion.


\noindent{\bf{Word Count}}\hspace{.1in}
\textit{Word count}~\cite{blumenstock2008size,pennebaker2001linguistic,ahmad2012puma} is a basic algorithm in
document analytics, which is widely used in applications like document
classification, clustering, and theme identification. It
counts the total appearances of every word in a given dataset, which may
consist of a number of files. 
\begin{itemize}[noitemsep,nolistsep]
  \item Input: \{file1, file2, file3, file4, file5, ...\}
  \item Output: <word1, count1>, <word2, count2>, ...
\end{itemize}
 

\noindent{\bf{Inverted Index}}\hspace{.1in}
      \textit{Inverted index}~\cite{ahmad2012puma} builds word-to-file index for a document dataset. It is worth noting that in some implementations of \textit{inverted index}, some extra operations are involved (e.g., getting and storing the term frequency) during the construction of the word-to-file index~\cite{zobel2006inverted}. The implementation in our study is pure in computing only word-to-file index as the purpose is to demonstrate how TADOC can be used for various tasks. (We have a separate benchmark, \emph{term vector}, for computing term frequencies.)

\begin{itemize}[noitemsep,nolistsep]
  \item Input: \{file1, file2, file3, file4, file5, ...\}
  \item Output: <word1, <file1>{}>, <word2, <file13>{}>, ...
\end{itemize}


\noindent{\bf{Sequence Count}}\hspace{.1in}
\textit{Sequence count}~\cite{ahmad2012puma,zernik1991lexical,lebart1998classification} counts the number of
appearances of every $l$-word sequence in each file, where $l$ is an
integer greater than 1.  In this work, we use $l$=3 as an
example. \textit{Sequence count} is very useful in semantic, expression, and
sequence analysis.  Compared to \textit{word count} and
\textit{inverted index}, \textit{sequence count} not only
distinguishes between different files, but also discerns the order of
consecutive words, which poses more challenges for processing (Section~\ref{sec:challenges}).

\begin{itemize}[noitemsep,nolistsep]
  \item Input: \{file1, file2, file3, file4, file5, ...\}
  \item Output: <word1\_word2\_word3, file1, count1>, ...
\end{itemize}

\vspace{-0.20in}
\section{TADOC and its Challenges}
\label{sec:challenges}
\vspace{-0.10in}

In this section, we present the concept of TADOC,
including its basic algorithms and the
challenges for materializing it effectively.

\vspace{-0.20in}
\subsection{TADOC}
\label{sec:idea}
\vspace{-0.10in}
 
\cred{
TADOC is a technique that supports various text analytics tasks directly on compressed data without decompression.
The compression is not task-specific.
For example,
we compress text files using TADOC,
    and the compressed data can be used directly to support text analytics tasks such as \emph{word count} and \emph{inverted index}.
The basic concept of compression-based document analytics is to
leverage the compression results for \emph{direct processing}
while avoiding unnecessary repeated processing of repeated content in
the original data.
}

\begin{figure}[!h]
    \centering
    \includegraphics[width=1\linewidth]{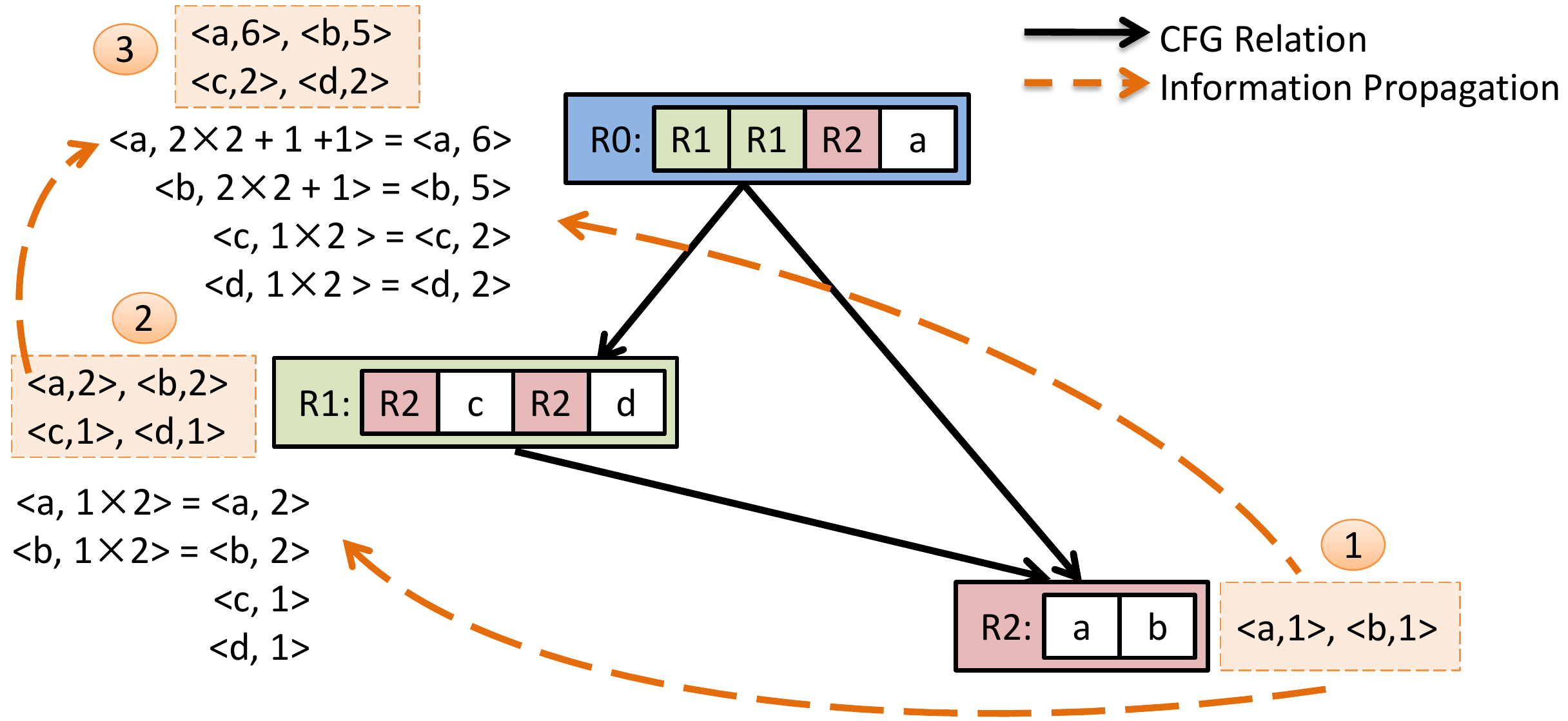}
      \caption{A DAG from Sequitur for ``a b c a b d a b c
        a b d a b a'', and a postorder traversal of the DAG for counting
        word frequencies.}\label{fig:wcExample}
\vspace{-0.20in}
\end{figure}

The results from Sequitur make this basic idea easy to materialize.
Consider a task for counting word frequencies in input
documents. We can do it directly on the DAG from Sequitur
using a postorder (children before parents) traversal,
as Figure~\ref{fig:wcExample} shows. After the DAG is loaded into memory, the traversal starts. At each node, it
counts the frequency of each word that the node directly
contains and calculates the frequencies of other words it indirectly
contains---in its children nodes. For instance,
when node R1 in Figure~\ref{fig:wcExample} is processed, direct
appearances of ``c'' and ``d'' on its right-hand-side (\textit{rhs}) are
counted, while, the frequencies of words ``a'' and ``b'' are
calculated by multiplying their frequencies in R2 by two---the number of
times R2 appears on its \textit{rhs}. When the traversal reaches the root R0,
the algorithm produces the final answer.


Because the processing leverages the
compression results, it naturally avoids repeated processing of
repeated content. 
The repeated content needs to be counted only once.
For instance, even though the substring ``a b'' (R2)
appears five times in the original string, the processing only counts
its frequency once. It is counted when the algorithm processes
R2; the results are reused for the two appearances of R2 when R1 is
processed; similarly, R1's results (and also R2's) are reused when R0
is processed.

\cred{
The procedure of postorder traversal for word counting is described in Algorithm~\ref{wordCountAlt1}, \texttt{wordcount-V1}.
The algorithm 
associates with each rule a
local table {\it locTbl}. During the data loading time (line \ref{wordCountAlt1:line:load} in
Algorithm~\ref{wordCountAlt1}), when reading in each rule in the CFG, the
algorithm records in the {\it locTbl} of that rule the frequency of
each rule and each word appearing on the right-hand side of that
rule. 
}
\vspace{-0.20in}

\begin{algorithm}[!h]
  \caption{\texttt{wordcount-V1}}
  \label{wordCountAlt1}
  \begin{algorithmic}[1]
    \algrenewcommand\textproc{}
    \State init()\Comment{nullify elements of Boolean array $done$}

    \State G=LoadMergeGraph(I)\Comment{Load compressed dataset $I$,
      build the graph with edges between two nodes merged into
      one; each node has a local table $locTbl$, initialized with the
      numbers of each subrule and each word's appearances on the
      right-hand side of the rule represented by that node.}
      {\label{wordCountAlt1:line:load}}
    \State countWords(G.root)

    \Procedure{countWords}{$id$}\Comment{$id$: a rule ID}
      \For{\texttt{each subRule $i$ in rule $id$}}
        \If {$!done[i]$}
          \State countWords\texttt{($i$)}
        \EndIf
      \EndFor

      \For{\texttt{each subRule $i$ in rule $id$}}
          \State $n$=$rule[id].locTbl[i]$\Comment freq of $i$ in rule $id$
          \For{\texttt{each word $w$ in the $locTbl$ of rule $i$}}
            \State {\texttt{$rule[id].locTbl[w]$ +=
                $n*rule[i].locTbl[w]$}}\Comment{fold the counts into
              the parent node}
          \EndFor
      \EndFor
      \State {$done[id] = true$}
    \EndProcedure
  \end{algorithmic}
\end{algorithm}

\vspace{-0.20in}
\cred{
  To analyze the complexity of Algorithm~\ref{wordCountAlt1},
    we first define four concepts in the DAG.
    }

\cred{
  \emph{Reachable}: In the DAG, if a path exists from node $r_i$ to node $r_j$, then we say node $r_j$ is reachable from $r_i$.
  }

\cred{
  \emph{Reachable node set}: In the DAG, 
    all the reachable nodes from node $r_i$ form a reachable node set of $r_i$.
  }

\cred{
  \emph{Reachable edge set}: In the DAG, 
    the out edges of all the nodes in node $r_i$'s reachable node set form the reachable edge set of node $r_i$.
  }
 
\cred{
  \emph{Reverse reachable node set}: After all the edges in the DAG are reversed, 
    all the reachable nodes from node $r_i$ form a reverse reachable node set of node $r_i$.
  }

\cred{
  \emph{Reverse reachable edge set}: After all the edges in the DAG are reversed, 
    the out edges of all the nodes in node $r_i$'s reachable node set form the reachable edge set of node $r_i$.
  }

\cred{
  The time complexity of Algorithm~\ref{wordCountAlt1} is dominated by two parts: 1) building a local table in each rule (lines 1-2),
    and 2) accumulating the word counts from the local table to the root (note that all the reverse reachable edges need to be traversed).
    For the first part,
    all the elements in each rule need to be analyzed,
    so the complexity involved in building the local table is $n_e\ast k$,
    where $n_e$ is the average number of elements for the nodes and $k$ is the number of nodes in the DAG.
    For the second part,
    assume the number of elements in the local word table of node $i$ is $n_i$ and the number of edges in the reverse reachable edge set of the node $i$ is $e_i$,
    then the complexity in accumulating the word counts from the node $i$ to the root is $n_i\ast e_i$.
    Based on the analysis,
    the time complexity of Algorithm~\ref{wordCountAlt1} is 
    $O(n_e\ast k+ \sum_{0}^{k-1}(n_i\ast e_i))$.
The space complexity of Algorithm~\ref{wordCountAlt1} is $O(s+g+k\ast l)$, where
$s$ is the size of the DAG, $g$ is the global table size, $k$ is the
number of nodes in the DAG, and $l$ is the average size of the local tables.
}

The example illustrates the essence of the general algorithm of our
TADOC method:

\noindent  \includegraphics[width=.5\textwidth]{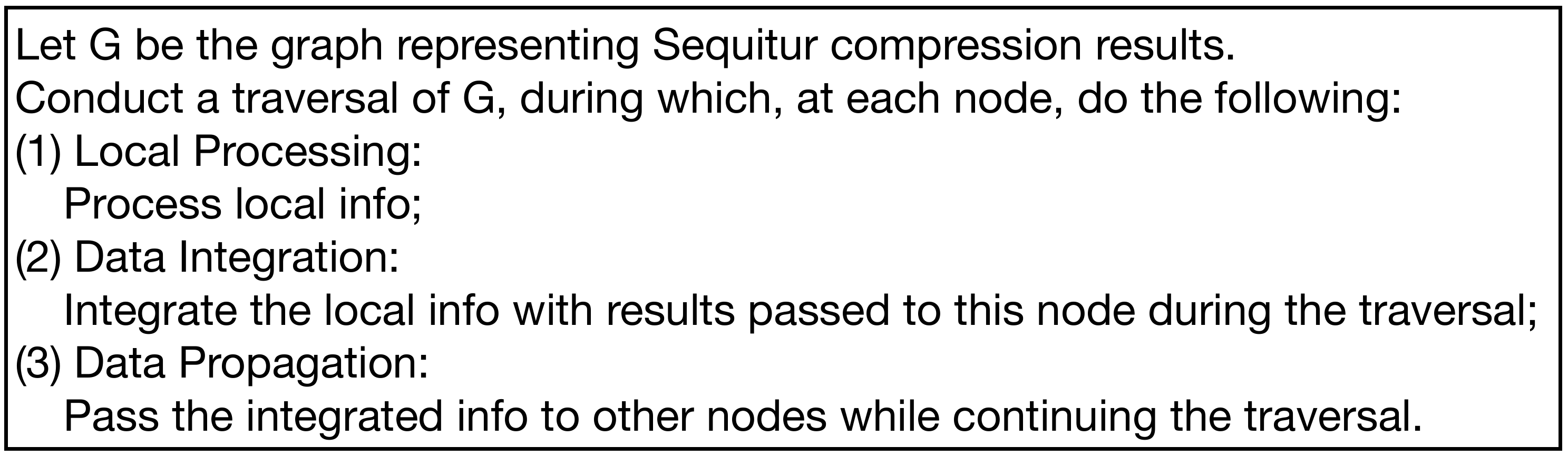}

We name the three main operations {\em local processing}, {\em data integration}, and {\em data propagation} respectively. 
Document analytics is converted into a graph
traversal process. 
Such a traversal process leverages the structure of the
input documents captured by Sequitur, and embodies information reuse
to avoid repeated processing of repeated content.







In terms of application scope, TADOC is
designed for document analytics applications that can be expressed as DAG
traversal-based problems on the compressed datasets, where the datasets
do \emph{not} change frequently.
Such applications would fit and benefit most from our approach.

\vspace{-0.20in}
\subsection{Challenges}
\label{sec:complexity}
\vspace{-0.10in}

\begin{figure}[!h]
\vspace{-0.15in}
  \center
  \includegraphics[width=.5\textwidth]{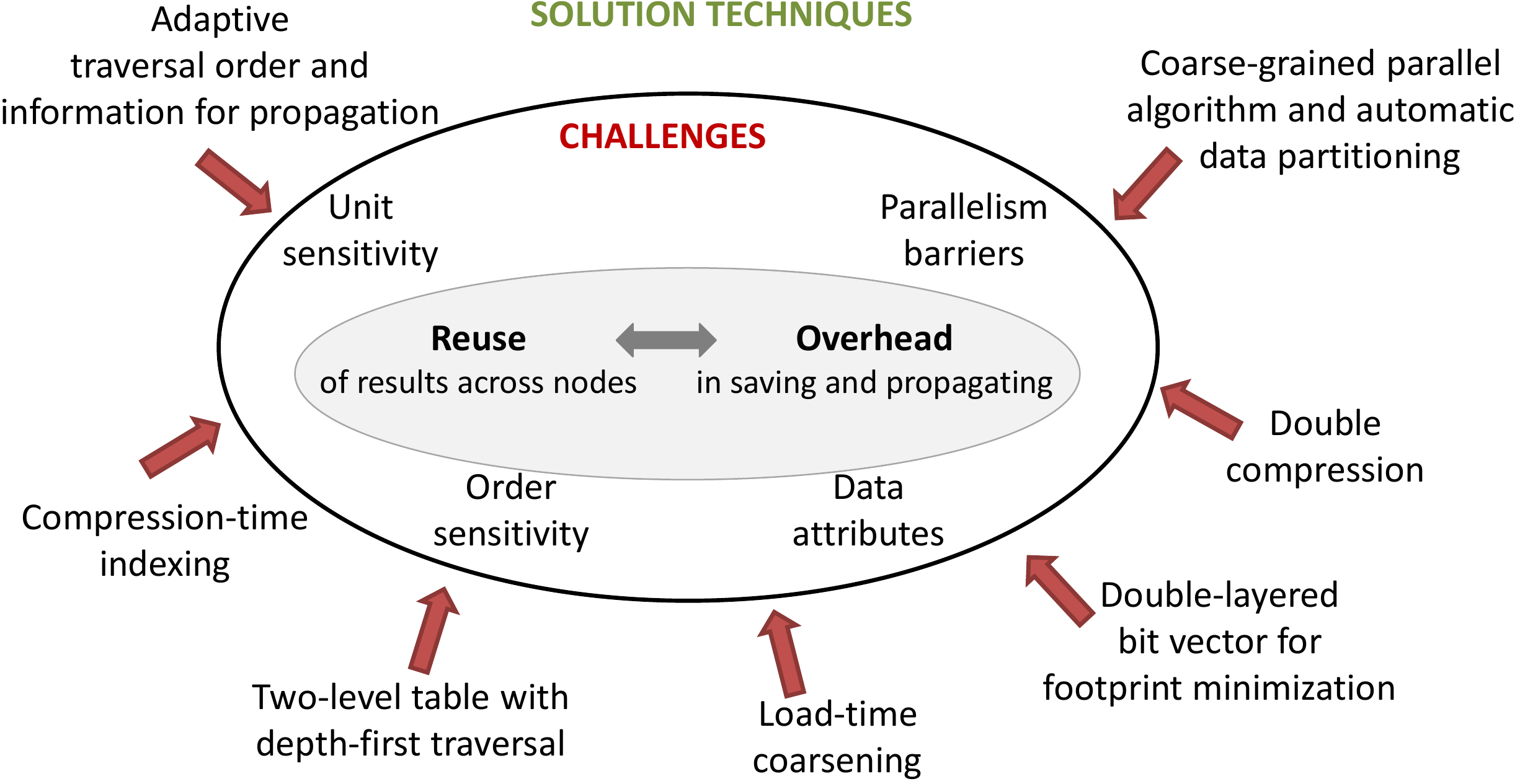}
\vspace{-0.20in}
  \caption{Overview of the challenges and solutions.}\label{fig:overview}
\vspace{-0.20in}
\end{figure}

Effectively materializing the concept of TADOC 
faces a number of
challenges. As Figure~\ref{fig:overview} shows, these challenges center around
the tension between reuse of results across nodes and the overheads
in saving and propagating results. Reuse saves repeated processing of
repeated content, but at the same time, requires the computation
results to be saved in memory and propagated throughout the graph. The
key to effective TADOC is to
maximize the reuse while minimizing the overhead.


Our problem is complicated by the complexities imposed by the various
analytics problems, the large and various datasets, and the demand for scalable
performance. We summarize the most common challenges as follows:

\begin{itemize}[noitemsep,nolistsep]
\item {\em Unit sensitivity}. \textit{Word count} regards the entire
  input dataset as a single bag of words. Yet, in many other document
  analytics tasks, the input dataset is regarded as a collection of
  some units (e.g., files). For instance, \textit{inverted index} and \textit{sequence count} try to get
  some statistics in each file. The default Sequitur compression does
    \emph{not} discern among files. How to support unit sensitivity is a question to be answered.
  

\item {\em Order sensitivity}. The way that results are propagated in the
  example of \textit{word count} in Figure~\ref{fig:wcExample}
  neglects the appearance order of words in the input documents. A challenge is how to
  accommodate the order for applications (e.g., \textit{sequence
    count}) that are sensitive to the order. This is especially tricky
  when a sequence of words span across the boundaries of multiple
  nodes (e.g., the ending substring ``a b a'' in
  Figure~\ref{fig:SequiturExample} spans across nodes R2 and R0).

\item {\em Data attributes}. The attributes of input datasets,
  such as the number of files, the sizes of files, and the number of unique words,
  may sometimes substantially affect the overhead and benefits of a
  particular design for TADOC.
    For instance, when solving \textit{inverted index}, one
  method is to propagate through the graph the list of files in which
  a word appears. This approach could be effective if there are a modest number of files,
  but would incur large propagation overheads otherwise,
    since the list to propagate could get very large. Thus, datasets
  with different properties could demand a different design in what to
  propagate and the overall traversal algorithm.


  
\item {\em Parallelism barriers.}  For large datasets, parallel
  or distributed processing is essential for performance.
 However, TADOC introduces
  some dependencies that form barriers. In Figure~\ref{fig:wcExample}, for instance, because
    nodes R1 and R0 require results from the processing of R2,
    it is difficult for them to be processed concurrently with R2. 



\end{itemize}


A naive solution to all these challenges is to decompress data before processing.
However, doing so loses most  benefits of compression. 
We next present our novel solutions to the challenges. 


\vspace{-0.25in}
\section{Guidelines and Techniques}
\label{sec:solutions}
\vspace{-0.15in}

This section presents our guidelines, techniques, and software
modules for easing programmers' jobs in implementing efficient TADOC.

\vspace{-0.25in}
\subsection{Solution Overview}
\vspace{-0.15in}

The part outside the challenge circle in Figure~\ref{fig:overview}
gives an overview of the solutions to the challenges. 
Because of the close interplay between various
challenges, each of our solution techniques simultaneously relates
with multiple challenges. They all contribute to our central goal:
maximizing reuse while minimizing overhead.

The first solution technique is about the design of the graph
traversal algorithm, emphasizing the selection of the traversal order
and the information to propagate to adapt to different problems and
datasets (Section~4.2). The second is about data structure design, which is
especially useful for addressing unit sensitivity (Section~4.2). The third is on
overcoming the parallelism barriers through coarse-grained parallel
algorithm design and automatic data partition (Section~4.3). The fourth addresses
order sensitivity (Section~4.4). The other three are some general optimizations to
be applied at compression time and graph loading time, useful for both
reducing the processing overhead and maximizing the compression
ratio (Section~4.5). For these techniques, we have developed some software modules
to assist programmers in using the techniques.

In the rest of this section, we describe each of the techniques along with the corresponding modules.

\vspace{-0.25in}
\subsection{Adaptive Traversal Order}
\label{sec:order}
\vspace{-0.15in}

The first of the key insights we learned through our
explorations is that graph traversal order significantly affects the
efficiency of TADOC. Its influence
is coupled with the information that the algorithm propagates through the
graph during the processing. The appropriate traversal order choice depends on the
characteristics of both the problems and the datasets.


 In this part of paper, we first draw on \textit{word count} and
 \textit{inverted index} as two examples to explain this insight, and
 then present the derived guideline and a corresponding software module
 to assist developers. Through the way, we will also explain some basic
 operations needed to handle unit sensitivity of document analytics.

 \vspace*{.1in} 
 \noindent \textit{\cred{An Alternative Algorithm for \textit{Word Count}}} 

 \cred{
 Figure~\ref{fig:wcExample} has already shown how \textit{word
   count} can be done through a postorder traversal of the CFG. Doing so
 can yield a decent speedup by saving computation
 (e.g., 1.4X on a 2.1GB {\em Wikipedia} dataset~\cite{wikipedia}, which is dataset E in Section~\ref{subsec:evaluationEva}). 
 However, we find that if we change the
 traversal to preorder (parents before children), the speedups can 
 almost double. 
 }

\begin{figure}[!h]
   \centering
   \includegraphics[width=1\linewidth]{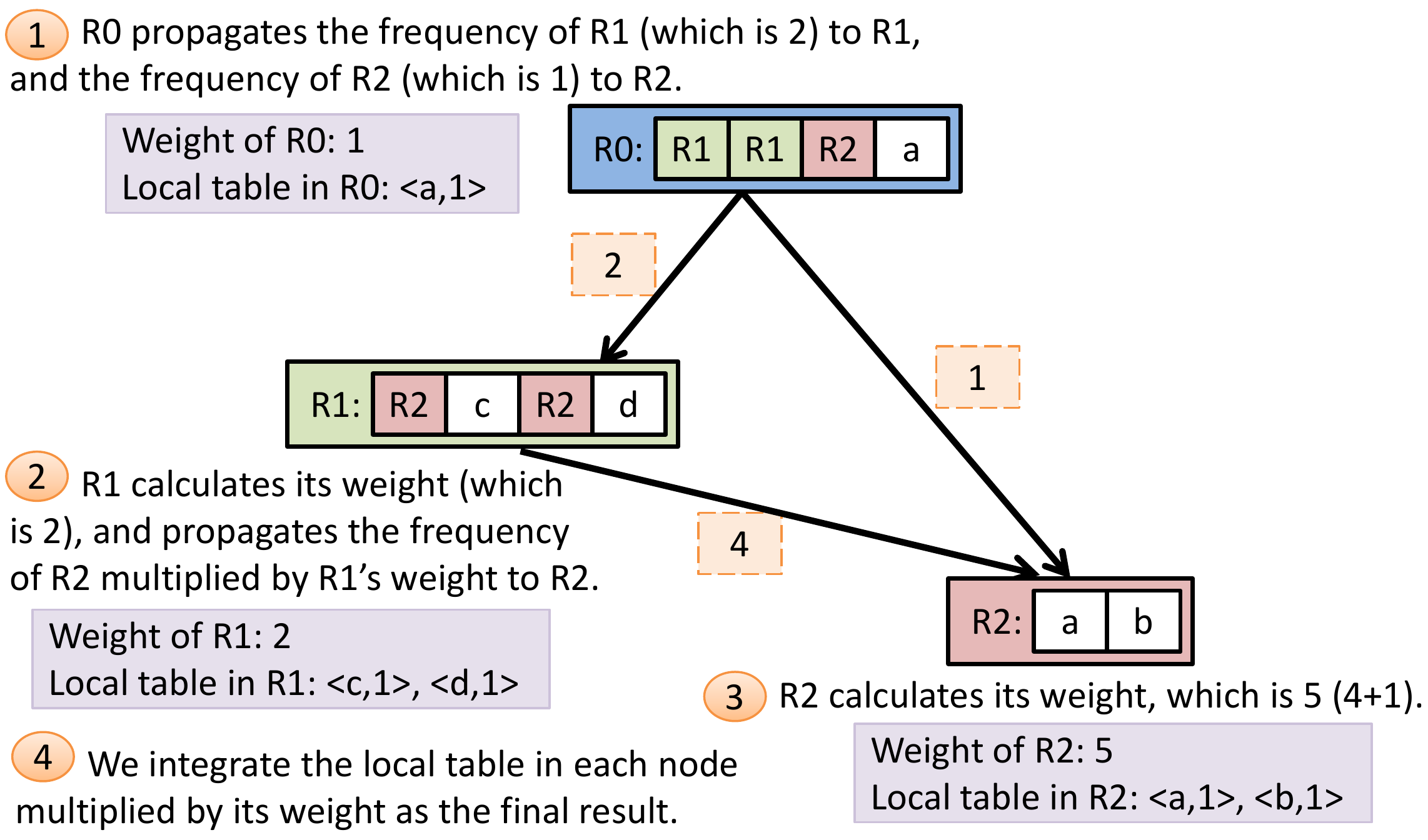}
\vspace{-0.20in}
     \caption{\cred{Illustration of a preorder traversal algorithm for
       \textit{word count} on the same string as
       Figure~\ref{fig:wcExample} deals with. The weights on an edge
       between two nodes (say, $node_x$ to $node_y$) indicate the total
       appearances of $node_y$'s rule due to all
       appearances of $node_x$'s rule.}}\label{fig:wcExample2}
\vspace{-0.20in}
 \end{figure}

 \cred{
Figure~\ref{fig:wcExample2} illustrates this alternative design. It
 consists of two main steps. The first step calculates the total
 frequency with which each rule appears in the entire dataset (\circled{1}
 to \circled{3} in Figure~\ref{fig:wcExample2}).  This step is done
 through a preorder traversal of the graph: Parent nodes pass their
 total frequencies to their children, from which the children nodes
 can calculate their total frequencies. Let $f(r)$ be the computed frequency of rule $r$.
 With this step done, the second
 step (\circled{4} in Figure~\ref{fig:wcExample2}) just needs to enumerate
 all the rules once. No graph traversal is needed. When it processes rule $r$, it calculates $f_r(w)=c_r(w)\ast f(r)$, where, $c_r(w)$ is the directly observable
 frequency of the word $w$ on the right-hand side of rule $r$ (i.e.,
 without considering the subrules of $r$), 
 and $f_r(w)$ is the total directly observable frequency of $w$ 
 due to all occurrences of rule $r$. 
 Let $f(w)$ be the frequency of word $w$.
 The algorithm adds $f_r(w)$ into $f(w)$, the accumulated frequency of $w$. So when the enumeration of all rules is done, the total count of every word is produced.
 For example, $f_0(a)$ is 1, $f_1(a)$ is 0, and $f_2(a)$ is 5, so $f(a)$ is 6.
 }

 \cred{
Based on the idea, we implement \texttt{wordcount-V2}, shown in
Algorithm~\ref{wordCountv3}. It consists of two main steps. The first
step (lines 5 to 14) calculates the frequencies of each rule, and the
second step (lines 16 to 20) goes through the words on the right-hand
side of each rule to obtain the accumulated total frequency of each
word. The first step propagates the {\it fq} of each node to its children
in a preorder (parents before children) traversal of the graph. Note
that a node's {\it fq} should be propagated to its children only after its
{\it fq} is fully ready---that is, all its parents' {\it fq}s have been folded
into its {\it fq}. Algorithm~\ref{wordCountv3} uses a queue \textit{workQue} to
ensure that: a rule is enqueued only when its {\it fq} has been updated
$k$ times, where $k$ is the number of its parents (line 10 in
Algorithm~\ref{wordCountv3}).
Compared to \texttt{wordcount-V1}, this version needs to propagate
only an integer {\it fq} between nodes, significantly reducing the
runtime overhead. 
}

\begin{algorithm}
  \caption{\texttt{wordcount-V2}}
  \label{wordCountv3}
  \begin{algorithmic}[1]
    \algrenewcommand\textproc{}
        \State $G=LoadMergeGraph(I)$\Comment{same as in
          Algorithm~\ref{wordCountAlt1} including setting up {\it
            locTbl} for each rule; additionally, each rule has a field
          {\it fq} set to 0, $updates$ set to 0}
        \State $rule[G.root].fq=1$
        \State \texttt{init $workQue$ to empty}
        \State \texttt{$\mathit{workQue}$.enque($\mathit{G.root}$)}
        \While{\texttt{!$\mathit{workQue}$.empty()}}\Comment{Calculate
                 frequencies}
        \State \texttt{$\mathit{head}$ = $\mathit{workQue}$.deque()}
        \For{\texttt{each subRule $i$ in rule $\mathit{head}$}}
        \State $rule[i].fq+=rule[head].fq\ast rule[head].locTbl[i]$
        \State {\texttt{$\mathit{rule[i].updates}$++}}
        \If {$\mathit{rule[i].updates}$ ==
          $\mathit{rule[i].inEdges}$}\Comment{Done with updating its
          {\it fq}, ready to be propagated to its children}
        \State {\texttt{$\mathit{workQue}$.enque($\mathit{i}$)}}
        \EndIf
      \EndFor
      \EndWhile
      \State $init(wordCounts)$\Comment{all zeros}
    \For{\texttt{each rule $i$}}\Comment{accumulate word's frequency}
      \For{\texttt{each word $w$ in $\mathit{rule[i]}$}}
      \State {\texttt{$wordCounts[w]$ +=
          $rule[i].locTbl[w]\ast rule[i].fq$}}
      \EndFor
    \EndFor
  \end{algorithmic}
\end{algorithm}

\vspace{-0.15in}

\cred{
  The time complexity of Algorithm~\ref{wordCountv3} is dominated by two parts: 
  1) building a local table in each rule,
and 2) calculating the accumulated frequency for each rule.
  The first part is the same as the first part in the complexity analysis of Algorithm~\ref{wordCountAlt1}.
  For the second part,
  the rule frequency is transmitted from the root to the children, cumulatively;
  each edge is passed only once, so the complexity of this part is the number of edges $d_G$.
  Therefore, the time complexity of Algorithm~\ref{wordCountv3} is 
    $O(n_e\ast k+ d_G)$,
    which is much simpler than the complexity of Algorithm~\ref{wordCountAlt1}.
The space complexity of Algorithm~\ref{wordCountv3} is also $O(s+g+k\ast l)$, where
$s$ is the size of the DAG, $g$ is the global table size, $k$ is the
number of nodes in the DAG, and $l$ is the average size of the local tables.
}


 \cred{
 This alternative algorithm achieves a much larger speedup
 (2.0X versus 1.4X on dataset E of {\em Wikipedia} in Section~\ref{subsec:evaluationEva})
 than the postorder algorithm does. The reason is that it needs to
 propagate only an integer---the node's frequency---from a node to its
 children, while the postorder algorithm in Figure~\ref{fig:wcExample}
 needs to propagate the frequencies of all the words the node and the node's
 successors contain. This example illustrates the importance of
 traversal order and the information to propagate for the efficiency of
 TADOC.
 }

\vspace*{.1in} \textit{\cred{Traversals for \textit{Inverted Index}}}

\cred{
The appropriate traversal order depends on both the analytic tasks and
datasets. We illustrate this point by describing two
alternative traversal algorithms designed for \textit{inverted index}.
}

\begin{figure}[!h]
\vspace{-0.20in}
  \centering
  \includegraphics[width=.95\linewidth]{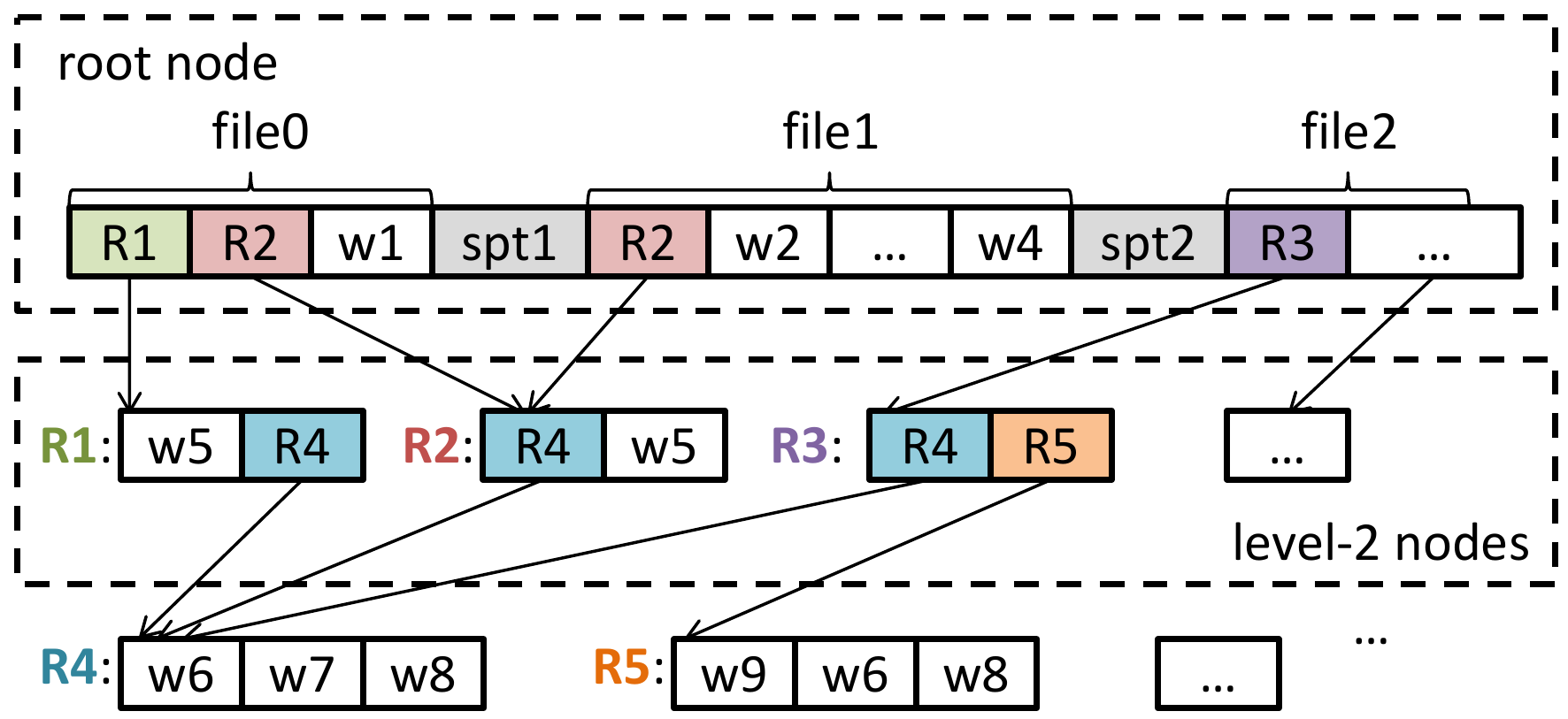}
  \caption{Sequitur compression result with file separators (``spt1'' and
    ``spt2'' are two file separators).
}\label{fig:indexing}
\vspace{-0.20in}
\end{figure}

Recall that the goal of \textit{inverted index} is to build a
mapping from each word to the list of files in which it
appears. Before we discuss the different traversal orders, 
we note that the objective of this analytics task 
requires discerning one file from another. 
Therefore, in the Sequitur
compression results, file boundaries should be marked.  
To do so, we
introduce a preprocessing step, which inserts some special markers at
file boundaries in the original input dataset. As these markers are
all distinctive and differ from the original data, in the Sequitur
compressed data, they become part of the root rule, separating the
different files, as the ``spt1'' and ``spt2'' in
Figure~\ref{fig:indexing} illustrate. (This usage of special markers
offers a general way to handle unit sensitivity.)

We next explain both preorder and postorder designs for \textit{inverted index}.
In the \emph{preorder} design, the first step propagates the
set of the IDs of the files that contain the string represented by the
node, instead of the frequencies of rules.  For instance, in
Figure~\ref{fig:indexing}, the {\it fileSet} of rule \texttt{R2} is
\{file0, file1\},
and rule \texttt{R3} is \{file2\}.
Because both rule \texttt{R2} and \texttt{R3} have rule \texttt{R4} as one of their subrules, during the
preorder graph traversal, the {\it fileSet} of rule \texttt{R4} is
updated to the union of their {\it fileSet}s as
\{file0, file1, file2\}. So, after the first step, every rule's {\it
  fileSet} consists of the IDs of all the files containing the
string represented by that rule. The second step goes through each
rule and builds the inverted indices by outputting the {\it fileSet}
of each word that appears on the right-hand side of that rule.

\cred{
The algorithm of preorder \textit{inverted index} is shown in Algorithm~\ref{invertedIdxExample1}.
Initialization part transmits the file information to the level-2 nodes, and places the level-2 nodes into the queue (lines~\ref{alg:inv1:a} to \ref{alg:inv1:b}).
After initialization, level-2 nodes are ready to propagate the file information to the nodes in lower levels.
Next, the algorithm propagates the file information so that each rule contains the file IDs it appears in
(lines~\ref{alg:inv1:c} to \ref{alg:inv1:d}).
It uses a queue:
each rule transmits its file IDs to its children (lines~9 to 13),
and places the children that finish gathering file IDs into the queue (lines~14 to 17).
Finally, the algorithm goes through the words and files for each rule to build the word-to-file index
(lines~\ref{alg:inv1:e} to \ref{alg:inv1:f}).
}

\begin{algorithm}
  \caption{\texttt{\cred{invertedindex-V1}}}
  \label{invertedIdxExample1}
  \begin{algorithmic}[1]
  \algrenewcommand\textproc{}
    \State $G=LoadMergeGraph(I)$
      \State \texttt{init $workQue$ to empty}
      \For{\texttt{each file $i$ in the root rule $0$}}  {\label{alg:inv1:a}}
        \State {\texttt{$initLevel2Nodes()$}}\Comment{Insert file info from the root and insert level-2 nodes into $workQue$}
      \EndFor  {\label{alg:inv1:b}}
    \While{\texttt{!$workQue$.empty()}}  {\label{alg:inv1:c}}
      \State {\texttt{$head$ = $workQue$.deque()}}
      \For{\texttt{each subRule $i$ in rule $head$}}
        \For{\texttt{each file $j$ in rule $i$}}
          \If {\texttt{$j$ not in $rule[i].fileSet$}}
              \State {$rule[i].fileSet$.insert($j$)}
          \EndIf
        \EndFor
        \State {\texttt{$rule[i].updates$++}}
        \If {$rule[i].updates == rule[i].inEdges$}
              \State {\texttt{$workQue$.enque(i)}}
        \EndIf
      \EndFor
    \EndWhile {\label{alg:inv1:d}}
     \For{\texttt{each rule $i$ in $\mathit{rules}$}}    {\label{alg:inv1:e}}
      \For{\texttt{each word $j$ in $\mathit{rule[i]}$}}
     \For{\texttt{each file $k$ in $\mathit{rule[i]}$}}
          \State {\texttt{insert $(j,k)$ into $result$}}
      \EndFor
    \EndFor   
    \EndFor    {\label{alg:inv1:f}}
  \end{algorithmic}
\end{algorithm}

\cred{
  The complexity of Algorithm~\ref{invertedIdxExample1} is dominated by two parts: 
  1) transmitting the file information from the root to all the children,
and 2) collecting the word-to-file relations from each rule to the global table.
For the first part,
in the root,
each rule element $h_i$ belongs to a file,
and the file information needs to be transmitted to all the reachable nodes of rule $h_i$.
Assuming the number of rule elements in the root is $h_G$,
and the number of the reachable edges of $h_i$ is $u_i$.
}
\cb{
Then, 
the time complexity of the first part is $O(\sum_{i=0}^{h_G-1}u_i)$.
For the second part,
assume that node $r_i$ belongs to $v_i$ files,
and the number of elements in node $r_i$ is $n_i$.
Then, the time complexity of the second part is 
 $O(\sum_{i=0}^{k-1}(v_i\ast n_i))$.
  Therefore, the time complexity of Algorithm~3 is 
    $O(\sum_{i=0}^{h_G-1}u_i+ \sum_{i=0}^{k-1}(v_i\ast n_i))$.
%
The space complexity of Algorithm~3 is also $O(s+g+k\ast l_{2lev})$, where
$s$ is the size of the DAG, $g$ is the global table size, $k$ is the
number of nodes in the DAG, and $l_{2lev}$ is the average size of the double-layered bitmaps (detailed in Section~\ref{sec:forCompression}).
}

The \emph{postorder} design recursively folds the set of words covered by a
node into the word sets of their parent node. The folding follows a
postorder traversal of the graph and stops at the immediate children
of the root node (called {\it level-2 nodes}.) The result is that
every level-2 node has a {\it wordSet} consisting of all the words
contained by the string represented by that node. From the root node,
it is easy to label every level-2 node with a {\it fileSet}---that is
the set of files that contain the node (and hence each word in its
{\it wordSet}). Going through all the level-2 nodes, the algorithm can
then easily output the list of containing files for each word, and
hence yield the inverted indices.

\cred{
  Algorithm~\ref{invertedIndexExample2} depicts postorder \textit{inverted index}.
  It consists of two main steps.
  The first step (lines \ref{alg:inv2:a} to \ref{alg:inv2:b}) performs word transmission in postorder to the level-2 nodes.
  Different from Algorithm~\ref{wordCountv3} of \texttt{wordcount-V2}, the algorithm only needs to record the word without its frequency.
  Because level-2 nodes contain all the word information needed,
  the second step only needs the root rule and the level-2 nodes to generate the inverted index (lines \ref{alg:inv2:c} to \ref{alg:inv2:d}).
}

\begin{algorithm}
  \caption{\texttt{\cred{invertedindex-V2}}}
  \label{invertedIndexExample2}
  \begin{algorithmic}[1]
\algrenewcommand\textproc{}

    \State $G=LoadMergeGraph(I)$
      \For{\texttt{each subrule $i$ in the root rule $0$}} {\label{alg:inv2:a}}
        \State {postorderTraverse($i$)}
      \EndFor  {\label{alg:inv2:b}}
      \For{\texttt{each file $i$ in the root rule $0$}}  {\label{alg:inv2:c}}
          \State {generateIndex($i$, $result$)}\Comment{Using root and level-2 nodes to generate the result}
      \EndFor  {\label{alg:inv2:d}}

    \Procedure{postorderTraverse}{$id$}
      \For{\texttt{each subRule $i$ in rule $id$}}
        \If {$!done[i]$}
          \State postorderTraverse\texttt{($i$)}
        \EndIf
      \EndFor

      \For{\texttt{each subRule $i$ in rule $id$}}
        \For{\texttt{each word $w$ in the $locTbl$ of rule $i$}}
          \State {$rule[id].locTbl[w] = true$}
          \EndFor
      \EndFor
      \State {$done[id] = true$}
    \EndProcedure
  \end{algorithmic}
\end{algorithm}

\cred{
  The complexity of Algorithm~\ref{invertedIndexExample2} is dominated by two parts: 
  1) transmitting the word set from the leaf to the level-2 nodes,
and 2) collecting the word-to-file relations from the root and the level-2 nodes.
For the first part,
the process is similar to Algorithm~1 and has the same complexity of 
    $O(n_e\ast k+ \sum_{0}^{k-1}(n_i\ast e_i))$,
    where $n_e$ is the average number of elements for the nodes, $k$ is the number of nodes,
    $n_i$ is the number of elements in the local word table of node $i$,
    and $e_i$ is the number of edges in the reverse reachable edge set of node $i$ in the DAG.
    For the second part,
    the number of words contained in each level-2 node depends on the input,
    and in the worst-case scenario, each level-2 node includes all vocabularies, whose number is $y$.
    Hence, the complexity of the second part is $h_G\ast y$,
    where  $h_G$ is the number of rule elements in the root.
    Therefore, the time complexity of Algorithm~4 is 
    $O(n_e\ast k+ \sum_{0}^{k-1}(n_i\ast e_i)+h_G\ast y)$.
The space complexity of Algorithm~4 is $O(s+g+k\ast l)$, where
$s$ is the size of the DAG, $g$ is the global table size, $k$ is the
number of nodes in the DAG, and $l$ is the average size of the local tables.
}

The relative performance of the two designs depends on the dataset.
For a dataset with many small files,
the preorder design tends to run much slower than postorder
(e.g., 1.2X versus 1.9X speedup over processing the original dataset directly on dataset D in
  Section~\ref{subsec:evaluationEva}, {\em NSF Research Award Abstracts}
 dataset~\cite{Lichman:2013}), because the file sets it propagates are large.
On the other hand, for a dataset with few large files,
the preorder design tends to be a better choice as the postorder design has to propagate large {\it wordSets}.

It is worth noting that \textit{word count} can also be implemented in both preorder and postorder, and preorder is a more efficient choice.

\vspace*{.1in} 
\noindent \textit{Guidelines and Software Module}

Our experience leads to the following two guidelines for implementing
TADOC.

{\bf Guideline I:} Try to minimize the footprint size of the data
propagated across the graph during processing. 

{\bf Guideline II:} Traversal order is essential for efficiency.
It should be selected to suit both the analytics task and the input
datasets.

These guidelines serve as principles for developers to follow
during their implementations of the solutions for their specific analytics tasks. 

Traversal order is worth further discussion. 
The execution time with
either order mainly consists of the computation time $\mathit{t_{compute}}$ and the data propagation time $\mathit{t_{copy}}$.
The former is determined by the operations performed on a node, while the latter by the amount of data propagated across nodes. Their values in a given traversal order are affected by both the analytics task and the datasets. Directly modeling $\mathit{t_{compute}}$ and $\mathit{t_{copy}}$ analytically is challenging.

We instead provide support to help users address the challenge through machine learning models.
For a given analytics problem, the developer may create multiple versions of the solution (e.g., of different traversal orders).
We use a decision tree model to select the most suitable version.
To build the model, we specify a list of features that potentially affect program performance.
According to the decision tree algorithm, these features are automatically selected and placed on the right place of the decision tree via the training process.
For training data,
we use some small datasets that have similar characteristics to the target input.

We develop a software module, {\em OrderSelector},
which helps developers to build the decision tree for version selection.
The developer can then specify these versions in the configuration file of {\em OrderSelector} as candidates, and provide a list of representative inputs on which the program can run. 
They may also specify some (currently Python) scripts for collecting certain features of a dataset that are potentially relevant to the selection of the versions.
This step is optional as {\em OrderSelector} has a set of predefined data feature collection procedures, including, for instance, the size of an original dataset, the size of its Sequitur CFG, the number of unique words in a dataset, the number of rules, and the number of files.
These features are provided as metadata at the beginning of the Sequitur compressed data or its dictionary, taking virtually no time to read. 
With the configuration specified, {\em OrderSelector} runs all the versions on each of the representative input to collect their performance data (i.e., running time) and dataset features.
It then invokes an off-the-shelf decision tree construction tool (scikit-learn~\cite{pedregosa2011scikit})
on the data to construct a decision tree for version selection. The decision tree is then used in the production mode of {\em OrderSelector} to invoke the appropriate version for a given compressed dataset.

\begin{figure}[!h]
  \centering
  \includegraphics[width=.85\linewidth]{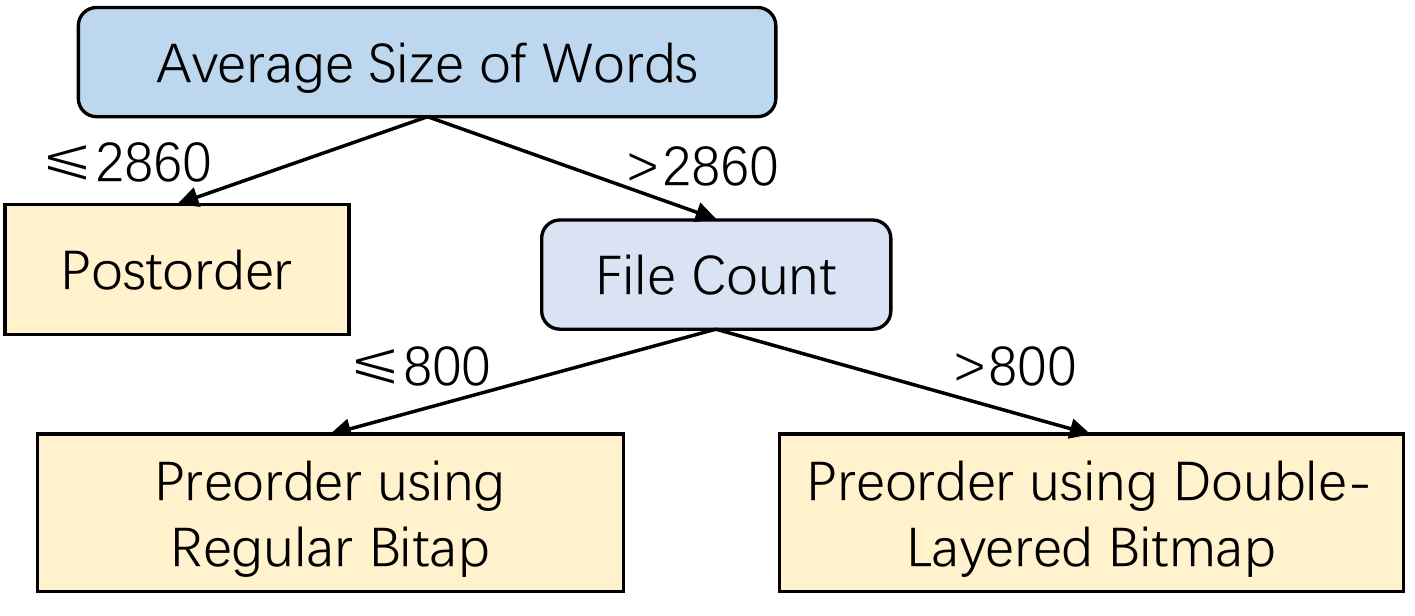}
\vspace{-0.10in}
  \caption{Decision tree for choosing traversal order.
}\label{fig:decisionTree}
\vspace{-0.30in}
\end{figure}

  Figure~\ref{fig:decisionTree} shows the decision tree obtained on {\em inverted index} based on the measurements of the executions of the different versions of the program on 60 datasets on the single node machine (Table~\ref{configurations}).
The datasets were formed by sampling the documents contained in the five datasets in Section~\ref{sec:evaluation}. They have various features: numbers of files range from 1 to 50,000, median file sizes range from 1KB to 554MB, and vocabulary sizes range from 213 to 3.3Million. The decision tree favors postorder traversal when the average file size is small (<2860 words) and preorder otherwise. (The two versions of preorder will be discussed in Section~\ref{sec:forCompression}). In five-fold cross validation, the decision tree predicts the suitable traversal order with a 90\% accuracy.

\subsection{Coarse-Grained Parallelism and Data Partitioning}
\label{sec:parallelism}
\vspace{-0.15in}

To obtain scalable performance, it is important for TADOC
to effectively leverage
parallel and distributed computing resources. As
Section~\ref{sec:challenges} mentions, some dependencies are introduced
between processing steps in either preorder or postorder traversals
of CFGs, which cause extra challenges for a parallel implementation.

We have explored two ways to handle such dependencies to parallelize the
processing.  The first is \emph{fine-grained partitioning}, which distributes
the nodes of the DAG to different threads, and inserts
fine-grained communication and synchronization among the threads to
exchange necessary data and results.  This method can potentially leverage existing work on parallel and distributed graph processing~\cite{gonzalez2012powergraph,petroni2015hdrf,carbone2015apache,martella2015practical,malewicz2010pregel,xin2013graphx}.
For instance, 
PowerGraph~\cite{gonzalez2012powergraph}, 
exploits the power-law property of graphs for distributed graph placement and representation,
and HDRF~\cite{petroni2015hdrf} is a streaming vertex-cut graph partitioning algorithm that considers vertex degree in placement.

The second is a \emph{coarse-grained partitioning} method. At compression time, this
method partitions the original input into a number of segments, then
performs compression and analytics on each segment in parallel,
and finally assembles the results if necessary.

The coarse-grained method may miss some compression opportunities
that exist across segments (e.g., one substring appears in two
segments). However, its coarse-grained partitioning helps avoid
frequent synchronization among threads. Our experimental results show
that on datasets of non-trivial sizes, the coarse-grained method
significantly outperforms the fine-grained method in both performance and
scalability. It is also easier to implement, for both parallel and
distributed environments. For a parallel system, the segments are
distributed to all cores evenly. For a distributed system, they are
distributed to all nodes evenly, and then distributed to the cores
within each node.

Load balance among threads or processes is essential for high
parallel or distributed performance.
Thus, the coarse-grained method requires
balanced partitioning of input
datasets among threads or processes. The partitioning can be done at the file
level, but it sometimes 
requires even \emph{finer} granularity
such that a
file is split into several sections, where each section is assigned
to a thread or process.

\vspace*{.1in} \noindent \textit{Guideline and Software Module}

Our experience leads to the following guideline. 

{\bf Guideline III:} 
Coarse-grained distributed implementation is preferred, especially when the input dataset exceeds the memory capacity of one machine; data partitioning for load balance should be considered, but with caution if it requires the split of a file, especially for unit-sensitive or order-sensitive tasks.





Dataset partitioning is important for balancing the load of the worker threads in coarse-grained parallelization.
Our partitioning mechanism tries to create subsets of files rather than splitting a file because there is extra cost for handling a split file, especially for unit-sensitive or order-sensitive tasks.  
To assist with this process, we develop a software module.  When the
module is invoked with the input dataset (a collection of files) and
the intended number of worker threads, it returns a set of partitions
and a metadata structure. 
  Resilient distributed dataset (RDD) is the basic fault-tolerant data unit in Spark~\cite{zaharia2010spark}.
The metadata structure records the mapping
relations among RDDs, files, and file sections. 
In the workload partitioning process, file splitting is considered only when a file exceeds a size threshold, $h_{split}$,
and causes significant load imbalance (making one partition exceed the average workload per worker by 1/4).
$h_{split}$ is defined as $\mathit{S_{total}}/2n_w$, where $\mathit{S_{total}}$ is the total dataset size, and $\mathit{n_w}$ is the number of workers.
The module 1) ensures that all workers process similar amounts of work
and 2) avoids generating small fragments of a file by
tolerating some disparity in the partition sizes. For applications that require
additional file or word sequence information, our partitioning
mechanism records some extra information, such as which file a section belongs
to, the sequence number of the section in the file, and so on. 
Such information is necessary for a thread to know
which section of which file it is processing, which is useful for
a later stage that merges the results.
\vspace{-0.30in}
\subsection{Handling Order Sensitivity}
\label{sec:treatment2orderSen}
\vspace{-0.10in}

As Section~\ref{sec:challenges} mentions, tasks that are sensitive to
the appearance order of words pose some special challenges. \textit{Sequence count}, for
instance, requires extra processing to 
handle 1) sequences that may
span across multiple Sequitur rules (i.e., nodes in the DAG) and 
2) order of words covered by different rules. The order sensitivity challenge (detailed in Section~3.2) 1) calls
for certain constraints on the visiting order of the rules in the
Sequitur grammar, and 2) demands the use of extra data structures to handle sequences across rules.

In our explorations, we found that the order sensitivity challenge can be addressed
through a two-level table design with a depth-first graph
traversal. The depth-first traversal of the graph ensures that the
processing of the data observes the appearing order of words in the
original dataset. During the traversal, we use a global table to
store the results that require cross-node examinations, and a
local table to record the results directly attainable from the right
hand side of the rule in a node. Such a design allows the visibility
of results across nodes, and at the same time, permits reuse of local
results if a rule appears many times in the dataset.

We take \textit{sequence count} as an example to illustrate our solution.
Algorithm~\ref{alg:seqCount} shows the pseudo-code.
The Sequitur design decides that the scanning process is similar to the depth-first graph traversal,
but the difference is that a sequence counting is integrated into the scanning process.
The general idea is to perform a depth-first traversal (line 5);
the word sequences that cross rules are stored in the global table,
while the word sequences within a rule are stored in the local table of each rule.
When the traversal is finished, we integrate the local tables from different rules to the global table (lines 6-7).
The \texttt{seqCount} function is used to process the rules, while the \texttt{process} function is used to process the words.
The depth-first graph traversal is embodied by the recursive function \texttt{seqCount} (lines 10 and 15 in Algorithm~\ref{alg:seqCount}).
It uses an $l$-element first-in first-out queue (\texttt{q}) to store the most-recently-seen $l$ words.
In function \texttt{process}, the most recent word is pushed into \texttt{q}, and this newly formed sequence in \texttt{q} is then processed,
resulting in an increment in the counters in either the local table \texttt{locTbl} (line 27) if the sequence does not span across rules, or otherwise, 
in the global table \texttt{gloTbl} (line 24).
The traversal may encounter a rule multiple times if the rule or its ancestors are referenced multiple times in the DAG. The Boolean variable \texttt{locTblReady} of a rule tracks whether the \texttt{locTbl} of the rule is ready to be reused, thus saving time.
  Note that in line 21,
  we also need to record the rule $r$,
  since we need to identify whether the words in $q$ are from multiple rules.

\begin{algorithm}
  \caption{Count $l$-word Sequences in Each File}
  \label{alg:seqCount}
  \small
  \begin{algorithmic}[1]
    \algrenewcommand\textproc{}
    \State $G=LoadData(I)$\Comment{load compressed data I; each rule
      has an empty $locTbl$ and a false boolean $locTblReady$}
    \State \texttt{allocate $gloTbl$ and an $l$-element long FIFO queue {\it q}}
    \For{\texttt{each file $f$}}
    \State $s=segment(f, G.root)$\Comment{Get a segment of the
      right-hand side of the root rule covering file $f$} (e.g., first three nodes in Figure~\ref{fig:indexing} for file0)
    \State $seqCount(s)$
    \State $calfq(s)$\Comment {calculate the frequency {\it fq} of each
      rule in $s$}
    \State $cmb(s)$\Comment{integrate into $gloTbl$ the $locTbl$ (times {\it fq}) of each rule subsumed by $s$}
    \State \texttt{output $gloTbl$ and reset it and {\it q}}
    \EndFor

    \Function{seqCount}{$s$}
    \For{\texttt{each element $e$ in $s$ from left to right}}
    \If {\texttt{$e$ is a word}}
    \State $process(e, s)$
    \Else
    \State $seqCount(e)$ \Comment{recursive call that materializes depth-first traversal of G}
    \EndIf
    \EndFor
    \State $s.locTblReady=true$
    \EndFunction

    \Function{process}{$e$, $r$}
    \State {\it q}$.enqueue(e, r)$\Comment{evict the oldest if full}
    \State \texttt{return if {\it q} is not full} \Comment{Need more words to form an $l$-element sequence}
    \If {\texttt{words in {\it q} are from multiple rules}}
    \State $gloTbl[\mathit{q.content}]++$
    \Else
    \If {!$r.locTblReady$}\Comment {avoid repeated processing}
    \State $r.locTbl[\mathit{q.content}]++$
    \EndIf
    \EndIf
    \EndFunction
  \end{algorithmic}
\end{algorithm}

\begin{figure}
\includegraphics[width=.98\linewidth]{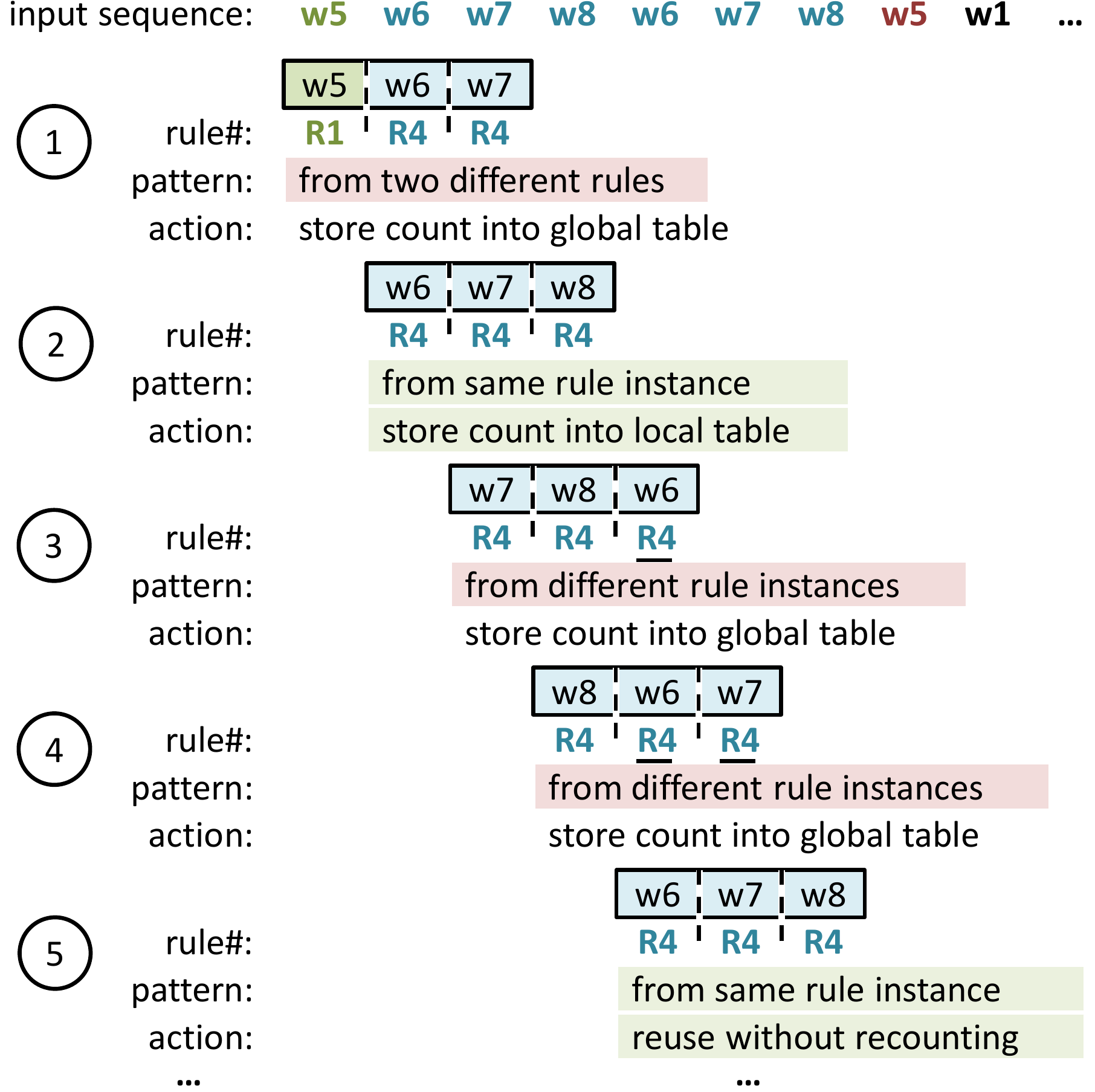}
  \centering
\vspace{-0.10in}
    \caption{Illustration of how Algorithm~\ref{alg:seqCount}
      processes an input sequence (DAG in Figure~\ref{fig:indexing})
      for counting 3-word long sequences.}
\label{fig:sequenceCountExa}
\vspace{-0.20in}
\end{figure}

Figure~\ref{fig:sequenceCountExa} demonstrates how
Algorithm~\ref{alg:seqCount} works on an input word sequence whose DAG
is shown in Figure~\ref{fig:indexing}. The words in the first
3-word sequence \circled{1} correspond to two different rules in the DAG (R1 and
R4). This sequence is a cross-node sequence and the algorithm stores its count
into a global table. The next 3-word sequence \circled{2} corresponds to only
R4, and is hence counted in a local table. The next two sequences \circled{3}, \circled{4} both
correspond to two instances of R4, and are both cross-node sequences.
Thus, they are counted in the global table. The bottom sequence \circled{5}
is the same as the second sequence \circled{2};
the algorithm does \emph{not} recount this
sequence, but directly reuses the entry in the local table of R4.

  The key for the correctness of Algorithm~\ref{alg:seqCount} is that the depth-first traversal visits \emph{all} words
  in the \emph{correct} order,
  i.e., the original appearance order of the words. We prove it as follows (``content of a node'' means the text that a node covers in the original document.)

\emph{Lemma 1}: If the content of every child node of a rule $r$ is visited in the correct order, the content of $r$ is visited in the correct order.

\emph{Proof:} Line 11 in Algorithm~\ref{alg:seqCount} ensures that the elements in rule $r$ are visited in the correct order. The condition of this lemma ensures that the content inside every element (if it is a rule) is also visited in the correct order. The correctness of the lemma hence follows.

\emph{Lemma 2}: The content of every leaf node is visited in the correct order.

\emph{Proof:} A leaf node contains no rules, only words. Line 11 in Algorithm~\ref{alg:seqCount} ensures the correct visiting order of the words it contains. The correctness hence follows.

\emph{Lemma 3}: The depth-first traversal by Algorithm~\ref{alg:seqCount} of a DAG G visits all words in an order consistent with the appearance order of the words in the original document G.

  \emph{Proof:} A basic property of a DAG is that it must have at least one topological ordering. In such an ordering, the starting node of every edge occurs earlier in the ordering than the ending node of the edge. Therefore, for an arbitrary node in G, its children nodes must appear \emph{after} that node in the topological ordering of G. Let $n_{-1}$ be the last non-leaf node in the ordering. \emph{Lemma 2} entails that all the nodes after $n_{-1}$ must be visited in the correct order as they are all leaf nodes, and \emph{Lemma 1} entails that the content of $n_{-1}$ must be visited in the correct order. The same logic leads to the same conclusion on the second to the last non-leaf node, then the third to the last, and so on, until the first node---that is, the root node. As the content of the root node is the entire document, \emph{Lemma 3} follows, by induction.

With \emph{Lemma 3}, it is easy to see that all $l$-long sequences in the original document goes through the FIFO queue \texttt{q} in Algorithm~\ref{alg:seqCount}. The algorithm uses the two tables \texttt{locTbl} and \texttt{gloTbl} to record the counts of every sequence in \texttt{q}. Function \texttt{cmb} folds all information together into the final counts. 

The computational complexity of Algorithm~\ref{alg:seqCount} depends mainly on two
functions, \texttt{seqCount} and \texttt{process}.  The complexity of
\texttt{seqCount} is determined by the total number of times rules are
visited, which is also the total number of times edges are traversed
in the DAG.
In reality, especially with coarsening to be described in
Section~\ref{sec:forCompression}, the overhead of \texttt{seqCount} is
much smaller than the overhead of \texttt{process}.  The complexity of
Algorithm~\ref{alg:seqCount} is practically dominated by
\texttt{process}, which has a complexity of $O(w)$.
$w$ is the number of words in the input documents.
The time savings
of Algorithm~\ref{alg:seqCount} over the baseline of direct processing on
the original data (i.e., without using our method) comes from avoiding repeatedly counting the sequences
that do \emph{not} span across rules. Thus, the amount of savings is
proportional to $m/n$, where $m$ is the number of repeated local
sequences and $n$ is the total number of sequences. The space
complexity of Algorithm~\ref{alg:seqCount} is $O(s+g+k\ast l)$, where,
$s$ is the size of the DAG, $g$ is the global table size, $k$ is the
number of nodes in the DAG, and $l$ is the average size of the local
tables.

\vspace*{.1in} \noindent \textit{Guideline}

{\bf Guideline IV:} For order-sensitive tasks, consider the use of
depth-first traversal and a two-level table design. The former helps
the system conform to the word appearance order, while the latter helps with
result reuse.

The global and local tables can be easily implemented through existing
template-based data structures in C++ or other languages. 
Hence, there is
no specific software module for the application of this guideline. 
The coarsening module
we describe next provides important performance benefits 
on top of this guideline.


\vspace{-0.20in}
\subsection{Other Implementation-Level Optimizations}
\label{sec:forCompression}
\vspace{-0.10in}

We introduce three extra optimizations. They are mostly implementation-level features that are useful for deploying TADOC efficiently. 




\vspace{1.0mm}
\noindent {\bf Double-layered bitmap}. As Guideline I says, minimizing the footprint of propagated data is
essential. In our study, we find that when dealing with unit-sensitive
analytics (e.g., \textit{inverted index} and
\textit{sequence count}), double-layered bitmap is often a
helpful data structure for reducing the footprint. 

\begin{figure}[!h]
\vspace{-0.18in}
 \centering
  \includegraphics[width=1\linewidth]{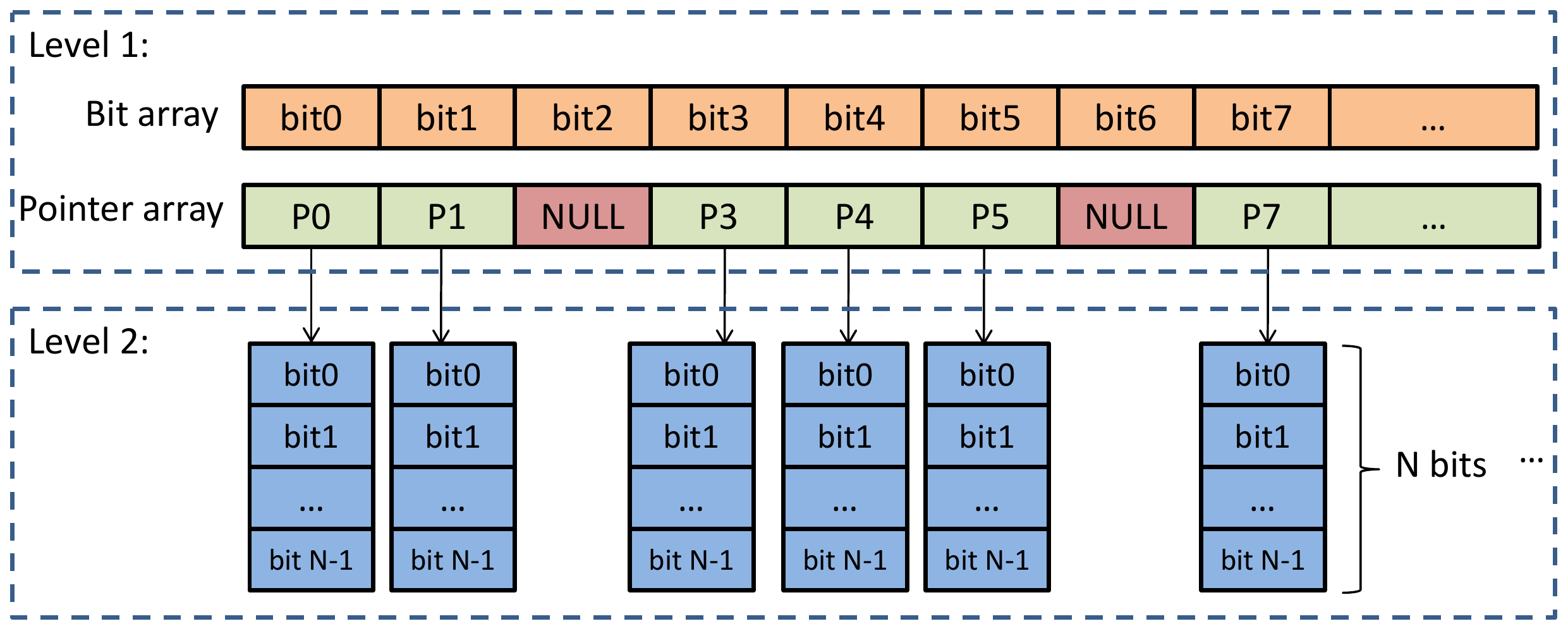}
\vspace{-0.20in}
  \caption{A double-layered bitmap for footprint minimization and
   access efficiency.}
  \label{fig:2levelBitMap}
\vspace{-0.28in}
\end{figure}

\cred{
Double-layered bitmap in this study is a data structure that includes level one and level two, and is used to store information about which files the rule belongs to in an efficient manner.
Level one is a data structure used to point to the locations for storing the file information in level-2 bitmaps for a given rule.
Level two is the data structure of bit vectors that actually store which files the rule belongs to.
In detail,
as Figure~\ref{fig:2levelBitMap} illustrates, level two contains a number of $N$-bit
vectors (where $N$ is a configurable parameter), while level one contains a
pointer array and a level-1 bit vector. The pointer array stores the
starting address of the level-2 bit vectors, while the level-1 bit
vector is used for fast checks to determine which bit vector in level
2 contains the bit corresponding to the file that is being queried. If a rule is
contained in only a subset of files whose bits correspond to some
bits in several level-2 vectors, the level two of the bitmap
associated with that rule would then contain only those several vectors,
and most elements in the pointer array would be null. The number of
elements in the first-level arrays and vectors of the double-layered
map is only $1/N$ of the number of files.  
For example,
assume that $N$ is four and there are 12 files.
The rule belongs to $file0$, $file1$, $file3$, $file4$, and $file5$.
Then, in level one,
$bit0$ and $bit1$ are \emph{true}, while $bit2$ is \emph{false};
$P0$ and $P1$ have pointer addresses and the value of $P2$ is \emph{NULL}.
In level two,
}
\cb{
the first bit vector is ``1101'' ($file0$, $file1$, and $file3$), the second bit vector is ``1100'' ($file4$ and $file5$),
and there is no bit vector that relates to $P2$.
}

Time-wise, the double-layered bitmap provides most of the efficiency
benefits of the one-level bit vector compared to the use of the set mechanism. Even
though it incurs one possible extra pointer access and one extra bit
operation compared to the use of one-level bit vectors, its total
memory footprint is much smaller, which contributes to better cache
and TLB performance. 
As Figure~\ref{fig:decisionTree} shows, the selection between the use of single-layer bitmap and double-layered bitmap can be part of the traversal order selection process. The decision tree in Figure~\ref{fig:decisionTree} favors double-layered bitmap when the average file size is greater than 2860 words and the number of files is greater than 800. 
(The benefits are confirmed experimentally in 
Section~\ref{sec:evaluation}).

It is easy to see that the double-layered bitmap can be used in
other applications with unit sensitivity as well. If the
unit is a class of articles, for instance, one just needs to change
the semantics of a bit to represent the class.

\noindent {\bf Double compression}. \textit{Double compression} is an optimization we have found helpful for the compression step. Compared with some other
compression algorithms, our Sequitur-based method is based on words and it does \emph{not}
always obtain the highest compression rates.
To solve this issue, we first compress the original dataset with
Sequitur and then run ``gzip''  on the output of Sequitur.
Note that ``gzip'' can be replaced with other methods with high compression rates, such as \texttt{zstd}~\cite{zstd} (\texttt{zstd} has a good compression ratio and is extremely fast at decompression).
    Considering the current universality that almost all platforms are equipped with \texttt{gzip},
    we shall still keep \texttt{gzip} as our default double-compression tool, but other compression tools are also supported in this work.
The result is often even more
compact than the ``gzip'' result on the original dataset.
To process the data,
one only needs to decompress the ``gzip'' result to recover the
Sequitur result. Because Sequitur result is usually much smaller than
the original dataset, it takes much less time to recover the Sequitur result than the
original dataset does. The decompression of the ``gzip'' result adds only a very small 
marginal overhead, as shown later.

\noindent {\bf Coarsening}. The third optimization relates to data loading time. It is called
{\em coarsening}, a transformation to the Sequitur DAG. Through it,
the nodes or edges in the graph can represent the accumulated
information of a set of nodes or edges. Specifically, we have explored
two coarsening methods: edge merging and node coarsening.  {\em Edge
  merging} merges the edges between two nodes in the DAG into one, and
uses a weight of the edge to indicate the number of original
edges. Merging loses the order of words, but helps reduce the size
of the graph and hence the number of memory accesses in the graph
traversal.  It is helpful to analytics tasks that are insensitive to
word order (e.g., \textit{word count} and \textit{inverted index}).
{\em Node coarsening} inlines the content of some small rules (which
represent short strings) into their parent rules; those small nodes
can then be removed from the graph.  It reduces the size of the graph,
and at the same time, reduces the number of substrings spanning across
nodes, which is a benefit especially important for analytics on word
sequences (e.g., \textit{sequence count}).  Coarsening adds some extra
operations, but the time overhead is negligible if it is performed
during the loading process of the DAG. On the other hand, it can save
memory usage and graph traversal time, as reported in the next section.

\vspace*{.05in} \noindent \textit{Guideline and Software Module}

{\bf Guideline V:} When dealing with analytics problems with unit
sensitivity, consider the use of double-layered bitmap if unit information
needs to be passed across the CFG. 

To simplify developers' job, we create a collection of double-layered
bitmap implementations in several commonly used languages (Java, C++,
C). Developers can reuse them by simply including the corresponding
header files in their applications.

Besides double-layered bitmap, another operation essential for handling unit sensitivity is the insertion of special markers into the documents to indicate unit boundaries when we do the compression as Section~\ref{sec:order} mentions.


{\bf Guideline VI:} 
{\em Double compression}
and {\em coarsening} help reduce space and time cost, especially when the dataset consists of many files.
They also enable that the thresholds be determined empirically (e.g., through decision trees).

We create two software modules to assist developers in using our
guideline. One module is a library function that takes original
dataset as input, and conducts Sequitur compression on it, during
which, it applies dictionary encoding and {\em double
  compression} automatically. In our implementation, this module and
the partitioning module mentioned in Section~\ref{sec:parallelism} are
combined into one compression module such that the input dataset first
gets integer indexed, then partitioned, and finally compressed. The
combination ensures that a single indexing dictionary is produced for
the entire dataset; the common dictionary for all partitions
simplifies the result merging process.

Our other module is a data loading module. When this module is invoked with
coarsening parameters (e.g., the minimum
length of a string a node can hold), it loads the input CFG with
coarsening automatically applied.

\vspace{-0.25in}
\subsection{Short Summary}
\vspace{-0.10in}

The six guidelines described in this section capture the most
important insights we have learned for unleashing the power of 
TADOC.
  They provide the solutions to
all the major challenges listed in Section~\ref{sec:complexity}:
Marker insertion described in Section~\ref{sec:order} and Guideline V together address unit sensitivity, Guideline IV order
sensitivity, Guideline II data attributes challenge, Guideline III
parallelism barriers, while Guidelines I and VI provide general
insights and common techniques on maximizing the efficiency. The
described software modules are developed to simplify the applications
of the guidelines. They form part of the \texttt{CompressDirect}
library, described next.


\vspace{-0.30in}
\section{C\MakeLowercase{ompress}D\MakeLowercase{irect} Library}
\label{sec:library}
\vspace{-0.10in}

We create a library named \texttt{CompressDirect} for two
purposes. The first is to ease programmers' burden in applying the six
guidelines when developing TADOC 
for an analytics problem. To this end, the first part of
\texttt{CompressDirect} is the collection of the software modules
described in the previous section. The second purpose is to provide a
collection of high performance implementations of some 
frequently-performed document analytics tasks, which can directly help many
existing applications.

Specifically, the second part of \texttt{CompressDirect} consists of
six high-performance modules. \texttt{Word
  count}~\cite{ahmad2012puma} counts the number of each word in all of the input
documents. \texttt{Sort}~\cite{huang2011hibench} sorts all the words
in the input documents in lexicographic order. \texttt{Inverted
  index}~\cite{ahmad2012puma} generates a word-to-document index
that provides the list of files containing each word. \texttt{Term
  vector}~\cite{ahmad2012puma} finds the most frequent words in a set
of documents. \texttt{Sequence count}~\cite{ahmad2012puma} calculates
the frequencies of each three-word sequence in every input file.
\texttt{Ranked inverted index}~\cite{ahmad2012puma} produces a list of
word sequences in decreasing order of their occurrences in each
document. These modules are essential for many text analytics applications. 

For each of these modules, we implement three versions: sequential,
parallel, and distributed. The first two versions are written
in C/C++ (with Pthreads~\cite{nichols1996pthreads} for
parallelization), and the third is in Scalar on
Spark~\cite{zaharia2010spark}. Our implementation leverages the
functions contained in the first part of the library,
which are the software modules described in Section~\ref{sec:solutions}.
A DAG is loaded into memory before it is processed. 
Large datasets are partitioned first with each partition generating a DAG that fits into the memory. The data structures used for processing are all in memory.
Using a Domain Specific Language may further ease the programming difficulty, as elaborated in a separate work~\cite{zhang2018zwift}. 
We next report the performance of our implementations.



\vspace{-0.25in}
\section{\cred{Supporting Advanced Document Analytics}}
\label{sec:advanced}
\vspace{-0.10in}

\cred{
In this section, we explore opportunities to apply TADOC to advanced document analytics.
We apply TADOC to three advanced document analytics applications: {\em word co-occurrence}~\cite{matsuo2004keyword,pennington2014glove},
\emph{term frequency-inverse document frequency} (TFIDF)~\cite{joachims1996probabilistic},
{\em word2vec}~\cite{rong2014word2vec,googleWord2Vec}, and {\em latent Dirichlet allocation} (LDA)~\cite{blei2003latent}.
}

%

\vspace{-0.25in}
\subsection{\cred{Word Co-Occurrence}}
\vspace{-0.10in}

\begin{figure}[!h]
  \center
\includegraphics[width=0.9\linewidth]{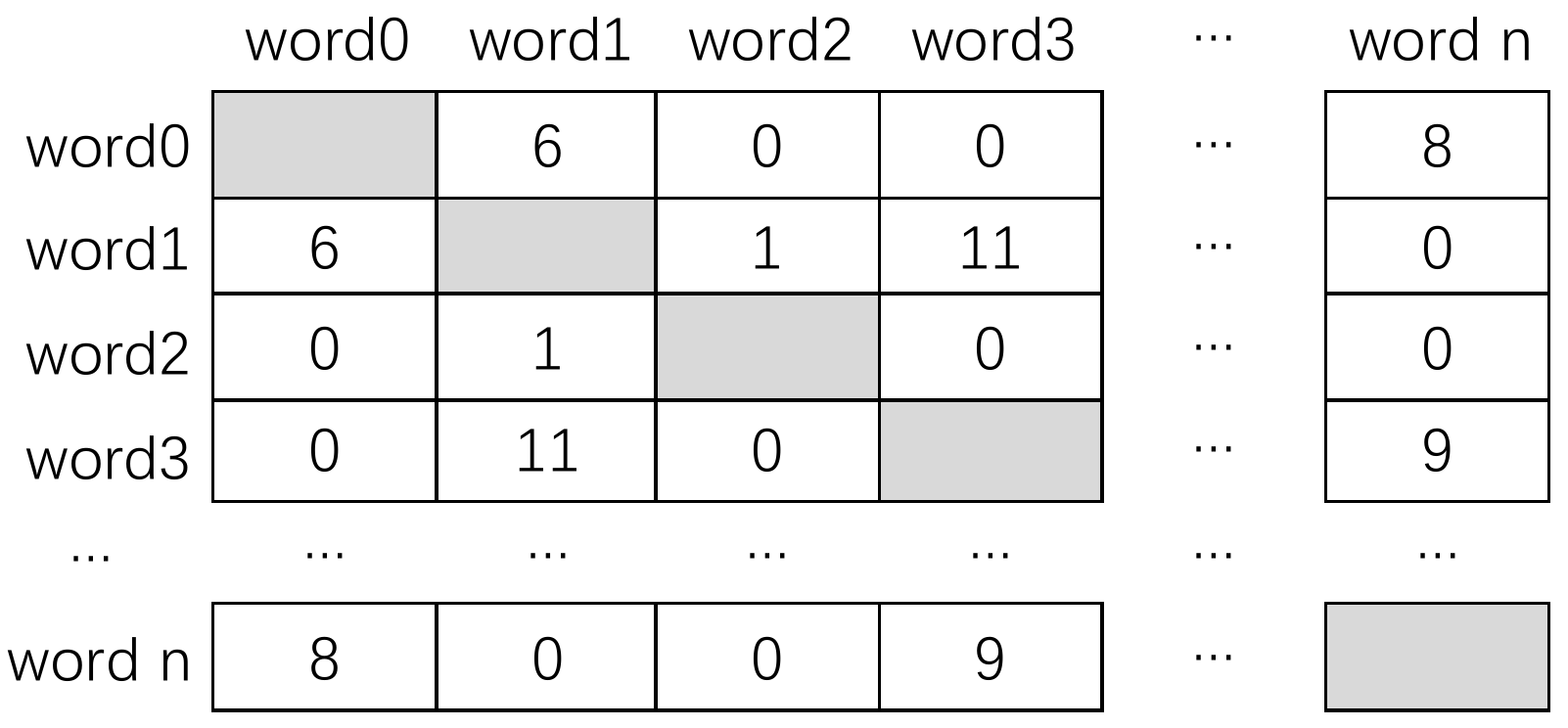}
\vspace{-0.10in}
  \caption{\cred{Word co-occurrence matrix example.}}
  \label{fig:wordcooccur}
\vspace{-0.23in}
\end{figure}

\cred{
Word co-occurrence is the frequency of the occurrence of two words alongside each other in a text corpus.
Because it can be regarded as a semantic proximity indicator,
word co-occurrence is widely used in linguistics, content analysis, text mining, and thesauri construction.
In general, word co-occurrence research aims to analyze similarities between word pairs and patterns,
and thus discovers latent linguistic patterns and structures in representations.
Word co-occurrence can be transformed into a word co-occurrence matrix, as shown in 
Figure~\ref{fig:wordcooccur}.
The rows and columns represent unique words,
and the numbers in the matrix denote the frequency at which $word_i$ co-occurs with $word_j$.
For example, in Figure~\ref{fig:wordcooccur},
$word_0$ co-occurs with $word_1$ six times
but does not co-occur with $word_2$.
}

\cred{
The TADOC technique can be applied as a data provider.
Because word co-occurrence only pertains to the frequency of adjacent words,
we can reuse the word co-occurrences in each rule.
A similar process also occurs in \emph{sequence count}.
Several optimizations can occur in this process.
In linguistics,
due to grammar and rules,
many words do not appear together;
hence, most of the elements in a matrix could be zero,
and the word co-occurrence data can be stored in a sparse matrix format, such as the compressed sparse row (CSR) and coordinate (COO) storage format~\cite{fengTKDE}.
In addition, a word co-occurrence matrix is a symmetric matrix when the word order is unnecessary,
in which case we store only the upper triangular part of the matrix.
We only need to coordinate our program interface with the corresponding word co-occurrence program interface.
}

\begin{figure}[!h]
\vspace{-0.20in}
  \center
\includegraphics[width=0.9\linewidth]{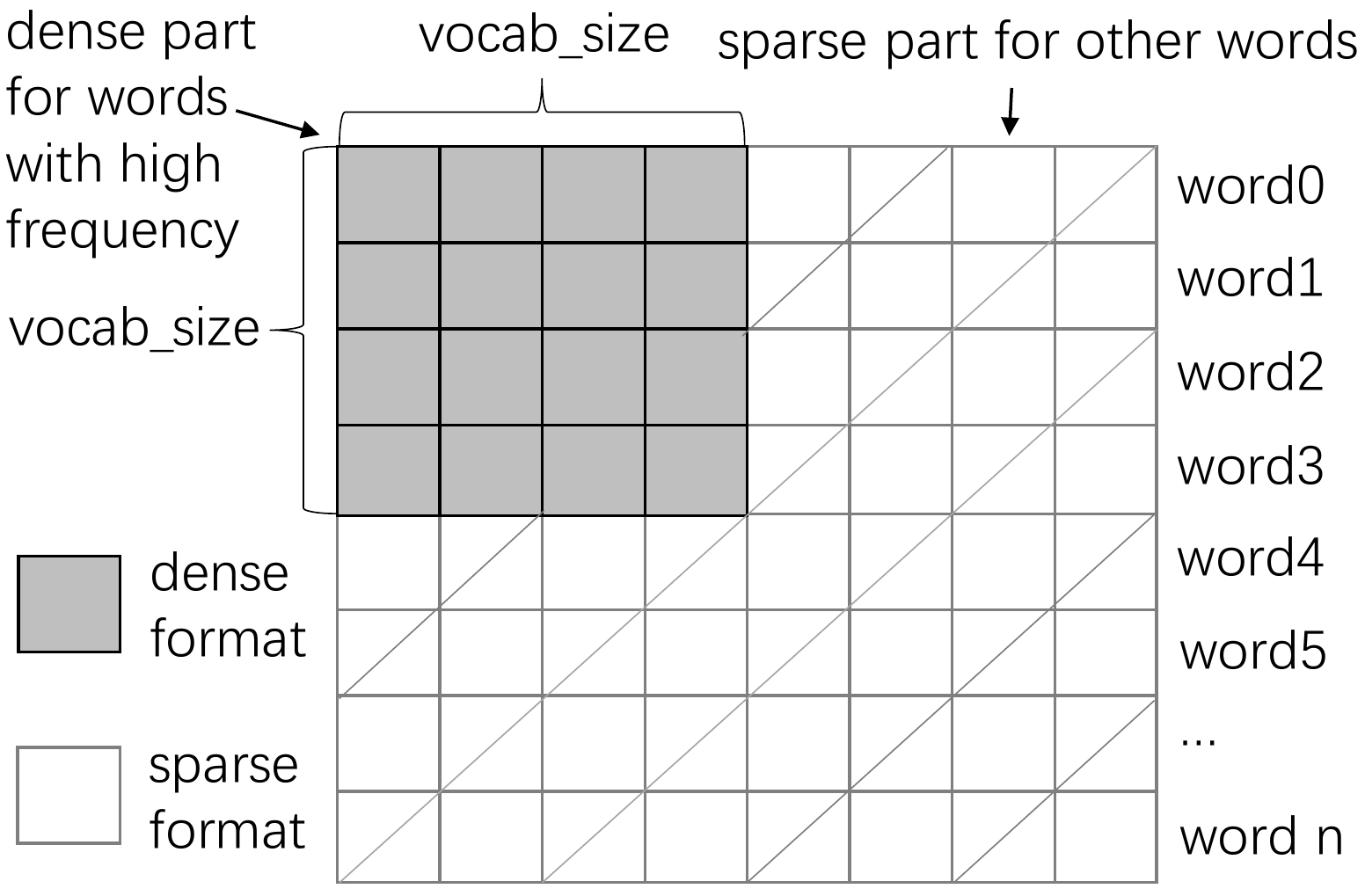}
\vspace{-0.10in}
  \caption{\cred{Word co-occurrence matrix in GloVe.}}
  \label{fig:glove}
\vspace{-0.20in}
\end{figure}

\cred{
  In this paper,
we use the word co-occurrence in GloVe~\cite{pennington2014glove} for validation.
The word co-occurrence matrix generated by this implementation is a mixture of dense and sparse matrices, as shown in Figure~\ref{fig:glove}.
The dense part represents frequent words that have a large number of occurrences;
these words are more likely to co-occur with other words,
and are therefore stored in a dense matrix.
The parameter $vocab\_size$ denotes the size of the dense matrix,
which can be adjusted by users.
The sparse part represents words that have a small number of occurrences,
and therefore uses a sparse matrix format.
Notably, GloVe uses a coordinate (COO) format for these sparse matrices.
}

\cred{
TADOC can be applied in the construction of dense and sparse matrices.
To separate words and build the two matrix parts,
the co-occurrence implementation first counts the frequency of each word,
and then calculates which words should be arranged in the dense matrix part.
The \texttt{CompressDirect} library directly provides word counts from the compressed format,
which is more efficient than providing these counts from the original uncompressed data.
Because the word count in \texttt{CompressDirect} reuses the redundant information,
it saves both computation and IO time.
}

\vspace{-0.20in}
\subsection{TFIDF}
\begin{figure}[!h]
  \center
\vspace{-0.20in}
\includegraphics[width=0.95\linewidth]{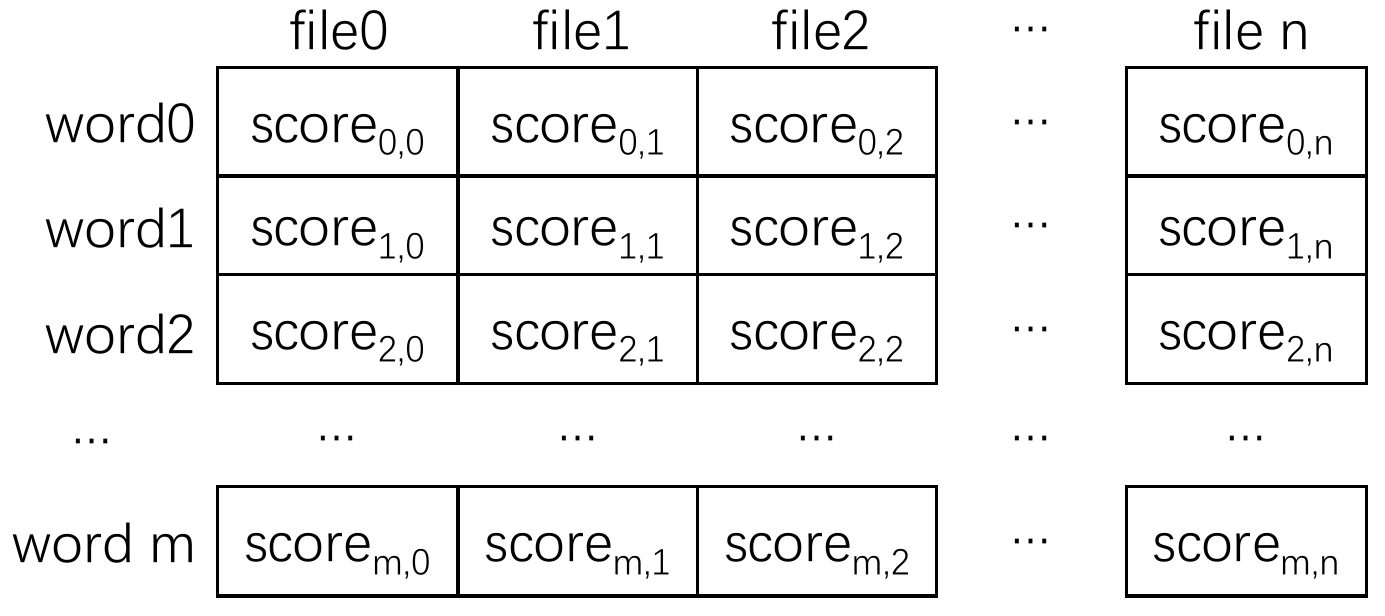}
\vspace{-0.10in}
  \caption{TFIDF model.}
  \label{fig:tfidf}
\vspace{-0.20in}
\end{figure}

\vspace{-0.10in}
\cred{
TFIDF~\cite{joachims1996probabilistic} is a statistical method used to evaluate the importance of a word to a document in the corpus.
The importance of a word increases proportionally with the frequency of the word in the document,
but it decreases inversely with the frequency of the word in the corpus.
TFIDF is a commonly used weighting technique for information retrieval and text mining,
and has been widely used in search engines as a measurement of the correlation between text and users.
Figure~\ref{fig:tfidf} shows an example of TFIDF;
$score_{i,j}$ exhibits the importance of $word_i$ to $document_j$.
}

\cred{
TFIDF consists of two parts: term frequency (TF), and inverse document frequency (IDF).
$TF(i,j)$ refers to the frequency of a given word $i$ that appears in document $j$.
$IDF_i$ is a measurement of the general importance of word $i$,
as shown in Equation~\ref{eq:idf}.
In Equation~\ref{eq:idf},
$\left | D \right |$ is the number of documents,
and $DF_i$ is the number of documents in which word $i$ appears at least once.
}\cb{
Given this definition of IDF,
}
\cred{
when a word occurs in a small number of files,
its importance is large; when the word occurs in many files,
its importance is small.
}

\begin{equation}\label{eq:idf}
  {
IDF_i=log(\frac{\left | D \right |}{DF_i})
}
\end{equation}

\cred{
In the TFIDF algorithm,
the $score_{i,j}$ of word $i$ to document $j$ is calculated using Equation~\ref{eq:tfidf}.
Intuitively,
word $i$ is important for document $j$ if the former occurs frequently in the latter,
but the importance of word $i$ decreases if the word occurs in many documents.
}

\begin{equation}\label{eq:tfidf}
  {
    score_{i,j}=TF(i,j)\cdot IDF_i
}
\end{equation}

    \cred{
TADOC can be used in the TFIDF algorithm.
The library 
\texttt{CompressDirect} generates the TF for each document,
and the TF generation process is similar to that of the \texttt{term vector}.
Because \texttt{CompressDirect} supports the \texttt{inverted index},
we can first execute the \texttt{inverted index} to calculate $DF$,
and then use the intermediate results to calculate the $IDF$ for each file.
}

\cred{
To validate the efficiency of TADOC on TFIDF,
we implement TFIDF~\cite{joachims1996probabilistic} using both the original data (baseline) and the compressed data.
The TFIDF algorithm can be divided into two stages.
In the first stage, TF and IDF are calculated.
Specifically,
we calculate the word frequency in each document in the dataset to obtain TF;
meanwhile,
we record the inverted word-to-document index,
which can be used to obtain IDF.
In the second stage, the TFIDF values are calculated using both TF and IDF,
which are represented as the scores in Figure~\ref{fig:tfidf}.
}

\cred{
The difference between the use of TFIDF with and without TADOC lies in the first stage.
In the baseline,
we need to process words sequentially in the original files to generate TF and IDF.
In the \texttt{CompressDirect} version,
the word frequencies and inverted index of each file are obtained by traversing the DAG,
during which deduplication occurs in the reuse of rules,
thus saving both space and computation time.
The second stages of the baseline and the \texttt{CompressDirect} version are the same.
}

\vspace{-0.20in}
\subsection{Word2vec}
\begin{figure}[!h]
  \center
\includegraphics[width=1\linewidth]{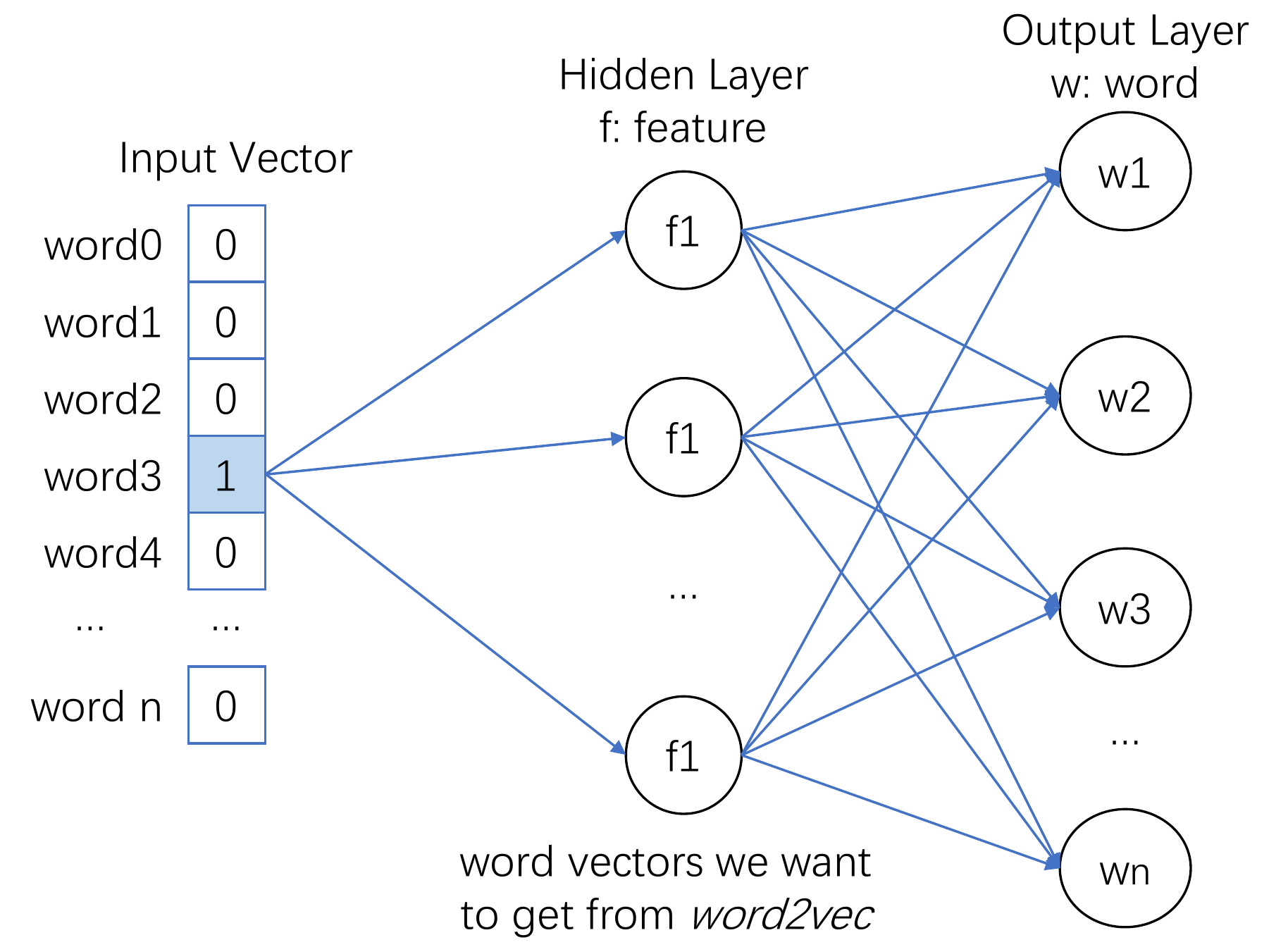}
\vspace{-0.20in}
  \caption{Word2vec example.}
  \label{fig:word2vec}
\vspace{-0.20in}
\end{figure}

\vspace{-0.10in}
\cred{
Word2vec~\cite{rong2014word2vec} is a shallow two-layer neural network that converts words from documents to feature vectors, as shown in Figure~\ref{fig:word2vec}.
The input of word2vec is a bag of text documents, and its outputs are feature vectors used to describe the words in the text corpus.
The vector output of word2vec can be used as input in many applications, such as long short-term memory (LSTM)~\cite{chiu2016named} and recommendation systems~\cite{vasile2016meta},
since this output represents words in a limited number of feature dimensions by numbers.
Originally, word2vec was designed to process text data,
but its applicability has been extended beyond the text corpus scope;
word2vec has applications in non-text patterns such as genes~\cite{koiwa2017extraction}, code~\cite{popov2017malware}, and social media~\cite{xu2017new}.
In this work, we mainly concentrate on the text document corpus.
}

\cred{
The goal of word2vec is to use a vector to represent each word;
the dimension of the vector is limited, but the vector can represent the meaning of the word with precise properties.
As shown in Figure~\ref{fig:word2vec},
the input to the network is a one-hot encoded vector for a word; the length of the vector is equal to the number of unique words,
and the vector of the index represents the corresponding word in the vocabulary.
The neurons in the hidden layer represent different features, so the neuron size is equal to the size of the word vector.
The output layer can be regarded as a probability vector;
each element represents the probability that a word appears around the input word;
hence, the probability vector length is equal to the vocabulary size.
Notably, once the training has finished,
we only need the word vector trained in the hidden layer, 
and the neuron network itself is useless;
further studies can be conducted with the generated word vectors.
}

\cred{
TADOC can be applied in the construction of the input vector.
Building the input vector from the compressed data with \texttt{CompressDirect} is more efficient than building the input vector from the original data,
not only due to the smaller storage size.
In real word2vec implementations,
the input vector is more complex than the input vector illustration in Figure~\ref{fig:word2vec},
and words need to be encoded according to different specifications.
Those coding specifications are usually related to word frequency,
and TADOC can efficiently provide the required word frequencies.
}



\cred{
We use the word2vec in~\cite{googleWord2Vec} for validation.
In this implementation,
before training,
word2vec has a preprocessing step,
during which word2vec
loads input data into the system and then encodes words.
In the encoding process,
Huffman coding~\cite{sharma2010compression} is used for the words;
first,  words are sorted according to their frequency,
and then shorter codes are assigned to words with higher frequencies.
In the next training step,
Huffman codes are used instead of the words.
As for TADOC,
Sequitur-compressed data can be used as input in the preprocessing step of word2vec.
Because word2vec needs the word frequency for Huffman coding,
the \texttt{CompressDirect} library provides the word counts in addition to the word sequence.
}

\vspace{-0.20in}
\subsection{Latent Dirichlet Allocation (LDA)}
\label{sec:lda}
\begin{figure}[!h]
\vspace{-0.20in}
  \center
\includegraphics[width=0.9\linewidth]{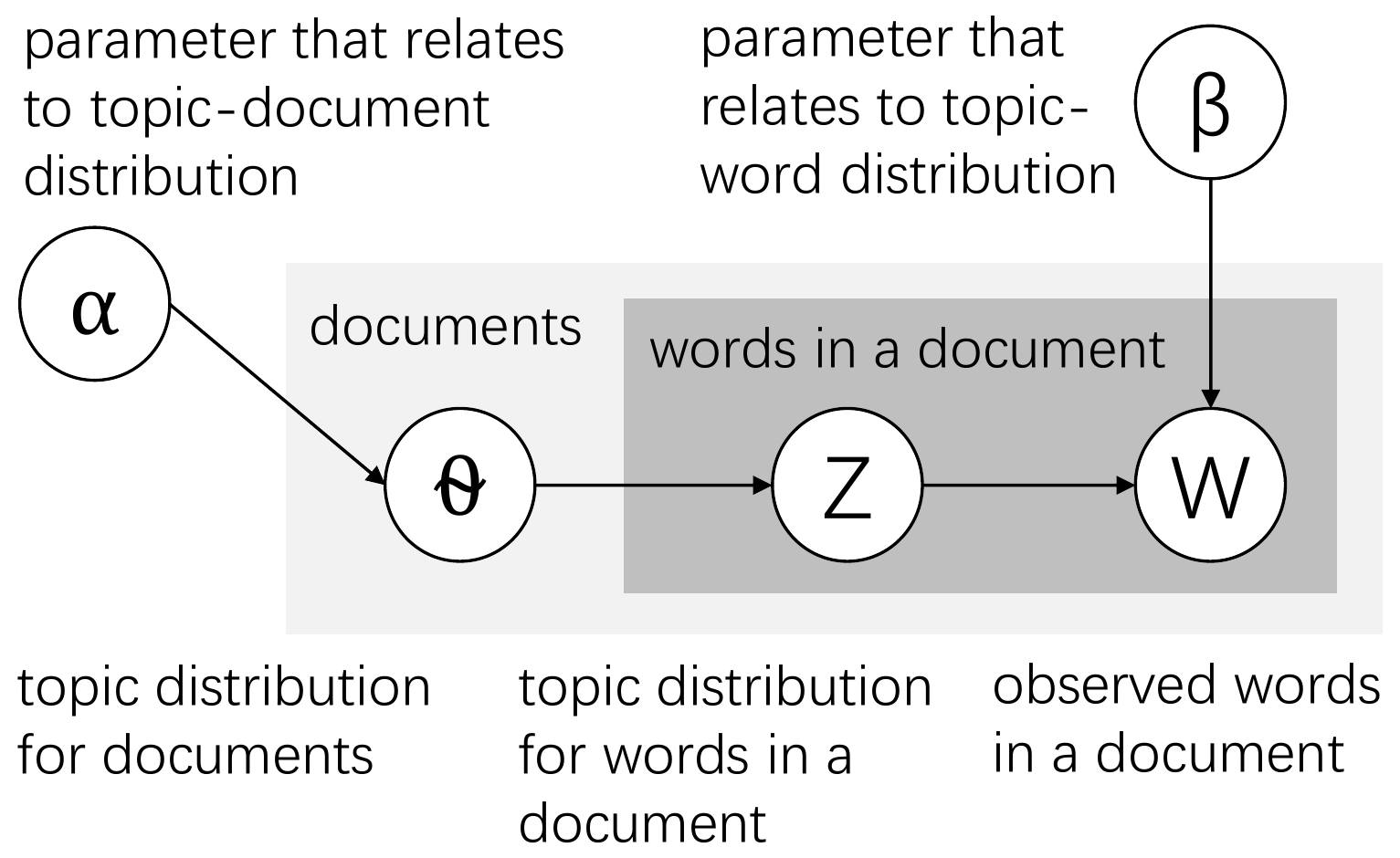}
\vspace{-0.10in}
  \caption{LDA model. The parameters $\alpha$ and $\beta$ usually have default values,
  so only the documents and the number of specified topics are input to LDA.}
  \label{fig:lda}
\vspace{-0.20in}
\end{figure}

\vspace{-0.10in}
\cred{
LDA~\cite{blei2003latent} is a generative probabilistic model
that is popular in machine learning and natural language processing,
where it is used to provide the subjects of documents in a probability form.
LDA is an unsupervised learning algorithm;
it does not require manual labeling.
The only information that users need to provide is the number of specified topics for the given documents.
In addition,
for each topic, LDA can provide relevant words (these words define the abstracted topic).
The graphical model representation of LDA is shown in Figure~\ref{fig:lda}.
}

\cred{
LDA assumes that documents are generated by latent topics, and each topic is characterized by a word distribution,
which explains why a bag of words appears in a given document.
The LDA model consists of five major components.
The parameter $\alpha$ relates to the topic-document distribution;
the topic distribution for documents, $\theta$, follows a Dirichlet distribution with parameter $\alpha$.
The topic distribution for words, $Z$, follows a multinomial distribution of $\theta$.
The words observed in a document, $W$, are determined by both $Z$ and $\beta$.
In general, $\alpha$ and $\beta$ are corpus-level parameters,
$\theta$ is a document-level variable,
and $Z$ and $W$ are word-level variables that can be sampled in each document.
In the training process of LDA,
the posterior distribution of latent variables of $\theta$ and $Z$ are estimated given $W$, $\alpha$, and $\beta$.
}

\cred{
The compressed data using TADOC can be directly applied to the input preprocessing of LDA.
The inputs to LDA are isolated words.
LDA is based on the ``bag-of-words'' assumption, that is, the order of words in a document can be neglected.
TADOC generates word counts by efficiently traversing the DAG, which means that the required input data is provided in a very effective manner.
}

\cred{
We use the LDA in~\cite{plda} as the evaluation platform.
In this LDA implementation,
the training process consists of two stages.
The first stage is a preprocessing stage;
the original input needs to be preprocessed into a sparse representation.
The sparse representation stores each document in one line.
In each line, 
each word and its word count are stored sequentially as <$word_i$,$count_i$>.
The second stage is a training process.
This stage uses the preprocessed data in the first stage
and trains the probabilistic model mentioned in Section~\ref{sec:lda} to estimate parameters.
In addition,
the LDA implementation~\cite{plda} provides a parameter that controls the number of iterations in the training process;
in our evaluation, we set this parameter to ten.
}

\cred{
TADOC can be applied in the first stage.
The baseline version processes the original input data into the required sparse format,
while the \texttt{CompressDirect} version provides the word counts in the required format directly from the compressed data.
In addition, because LDA adopts a ``bag-of-words'' assumption in which word order does not need to be maintained,
we can simply use the word counts in \texttt{CompressDirect} in this stage.
For the second stage,
the baseline and the \texttt{CompressDirect} version have the same procedure.
}

\vspace{-0.20in}
\subsection{Other Advanced Document Analytics}
\vspace{-0.10in}
\cred{
TADOC can be applied to other advanced document analytics.
To realize the high-level goals of advanced document analytics such as LSTM,
the original raw text first needs to be converted into a vector format.
}\cb{
For this purpose, word2vec can be used with  TADOC, as we already demonstrated.
After the conversion, 
these vectors have high-level usage models,
}
\cred{
which are independent from our technique.
These independent techniques only use in-memory data structures without accessing the input raw data again,
which is orthogonal to the problem we are addressing.
Therefore, our technique can be used to support these advanced uses of document analytics, especially at the raw text processing stage.
}



\vspace{-0.20in}
\section{Evaluation}
\label{sec:evaluation}
\vspace{-0.10in}

Using the six algorithms listed at the end of the previous section, we
evaluate the efficacy of the proposed Sequitur-based document
analytics for both space and time savings. The baseline implementations
of the six algorithms come from existing public benchmark suites,
\texttt{Sort} from HiBench~\cite{huang2011hibench} and the rest
from Puma~\cite{ahmad2012puma}. We report performance in
both sequential and distributed environments. For a fair comparison, the
original and optimized versions are all ported to C++ for the
sequential experiments and to Spark
for the distributed experiments.

The benefits are
significant in both space savings and time savings. Compared to the
default direct data processing on {\em uncompressed} data, our method speeds
up the data processing by more than a factor of two in most cases, and
at the same time, saves the storage and memory space by a factor of 6
to 13. After first explaining the methodology of our experimental evaluation, we
next report the overall time and space savings, and then describe the
benefits coming from each of the major guidelines we have described earlier
in the paper. 


\vspace{-0.20in}
\subsection{Methodology}
\label{subsec:evaluationEva}
\vspace{-0.10in}


%

\noindent {\bf Evaluated Methods}
We evaluate three methods for each workload-dataset combination.
The ``baseline'' method processes the dataset directly, as
explained at the beginning of this section. The ``CD'' method is our
version using \textit{TADOC}. 
The input to ``CD'' is the dataset compressed using ``double compression'' (i.e., first compressed by Sequitur then compressed by Gzip).
The ``CD'' method first recovers the Sequitur compression result by undoing the Gzip compression,
and then processes the Sequitur-compressed data directly. 
The measured ``CD'' time covers all the operations.
The ``gzip'' method represents existing decompression-based methods. It uses Gzip
to compress the data.
At processing time, 
it recovers the original data and processes it.

\vspace{1.6mm}
\noindent \textbf{Datasets} We use five datasets for evaluations, shown in
Table~\ref{table:dataset}. They
consist of a range of real-world documents of varying lengths,
structures and content. 
The first three, \texttt{A, B, C},
are large datasets from Wikipedia~\cite{wikipedia},
used for tests on clusters. \texttt{Dataset D} is NSF
Research Award Abstracts (NSFRAA) from UCI Machine Learning
Repository~\cite{Lichman:2013}, consisting of a large number (134,631)
of small files.  \texttt{Dataset E} is a collection
of web documents downloaded from the Wikipedia database~\cite{wikipedia},
consisting of four large files.

\begin{table}[!h]
\vspace{-0.20in}
\centering
\caption{Datasets (``size'': original uncompressed size).}
\vspace{-0.10in}
\resizebox{\linewidth}{!}{
\begin{tabular}{lrrrr}
\hline

  Dataset    &  \multicolumn{1}{c}{Size}     & \multicolumn{1}{c}{File \#}  & \multicolumn{1}{c}{Rule \#}   & \multicolumn{1}{c}{Vocabulary Size}  \\\hline
A  & 50GB      & 109      & 57,394,616 & 99,239,057   \\ 
B  & 150GB     & 309      & 160,891,324 & 102,552,660    \\ 
C  & 300GB     & 618      & 321,935,239 & 102,552,660   \\
D  & 580MB     & 134,631 & 2,771,880 & 1,864,902   \\
E  & 2.1GB     & 4       & 2,095,573 & 6,370,437 \\\hline 
\end{tabular}
}
\label{table:dataset}
\vspace{-0.20in}
\end{table}

The sizes shown in Table~\ref{table:dataset} are the original dataset
sizes. They become about half as large after dictionary encoding (Section~\ref{subsec:sequitur}).
The data \emph{after} encoding is used for all experiments, including the baselines. 



\vspace{1.6mm}
\noindent \textbf{Platforms} The configurations of our experimental
platforms are listed in Table~\ref{configurations}. For the distributed experiments, we use
the \texttt{Spark Cluster}, a 10-node cluster on Amazon EC2~\cite{amazon2010amazon}, and the three
large datasets. The cluster is built
with an HDFS storage system~\cite{borthakur2008hdfs}. The Spark
version is 2.0.0 while the Hadoop version is 2.7.0. For the sequential experiments, we use the \texttt{Single Node} machine on the two smallest datasets.

\begin{table}[!h]
\centering
\caption{Experimental platform configurations.}
\vspace{-0.10in}
\resizebox{\linewidth}{!}{
\begin{tabular}{lll}
\hline
Platform       &   Spark Cluster& Single Node       \\\hline
OS             & Ubuntu 16.04.1 &Ubuntu 14.04.2  \\
GCC            &   5.4.0        &  4.8.2         \\
Node\#         &   10           & 1              \\
CPU            &Intel E5-2676v3 & Intel i7-4790  \\
Cores/Machine  & 2              & 4              \\
Frequency      &  2.40GHz       & 3.60GHz        \\
MemorySize/Machine & 8GB        & 16GB \\\hline

\end{tabular}
}
\label{configurations}
\end{table}



\vspace{-0.30in}
\subsection{Time Savings}
\label{sec:speedups}
\vspace{-0.10in}

\subsubsection{Overall Speedups}
\label{sec:speedupDistributed}
\vspace{-0.10in}

\begin{figure}[!h]
\vspace{-0.10in}
\centering
  \includegraphics[width=1\linewidth]{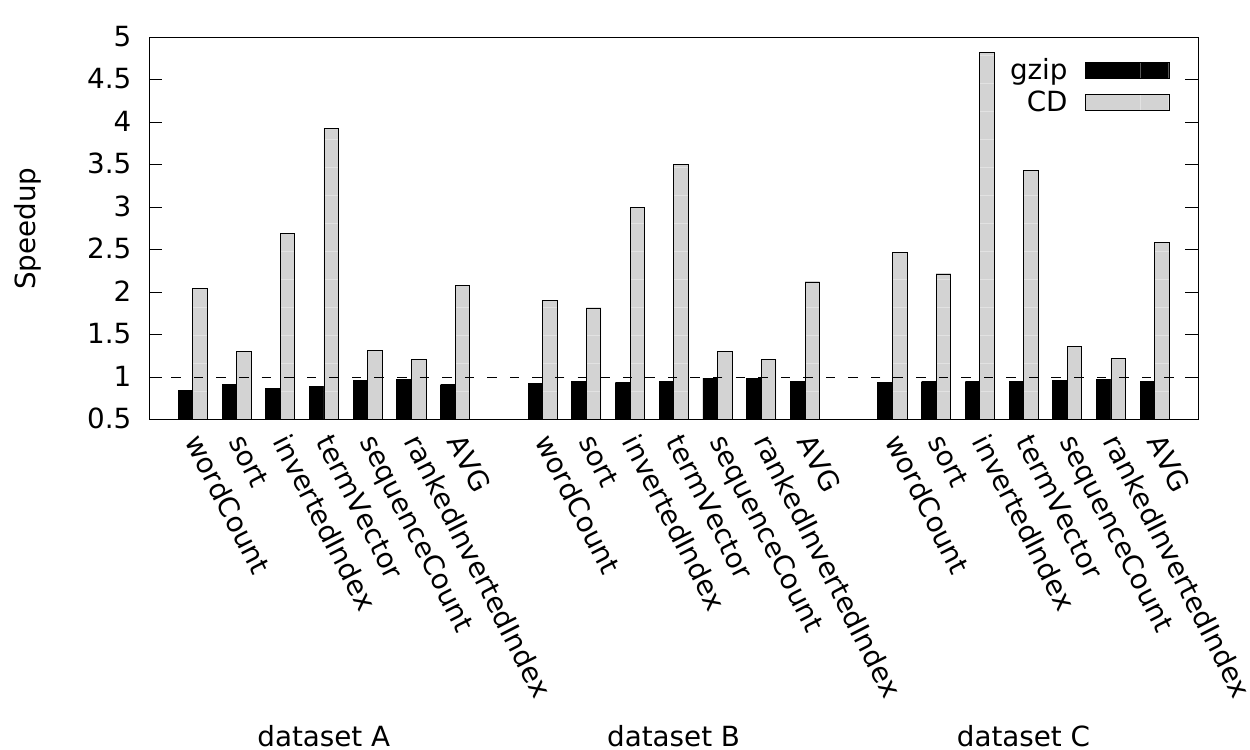}
\vspace{-0.15in}
\caption{Performance of different methods on large datasets running on the Spark
  Cluster, normalized to the performance of the baseline method.}
\label{fig:largerDataSets}
\vspace{-0.20in}
\end{figure}

Figure~\ref{fig:largerDataSets} reports the speedups that the different methods
obtain compared to the default method on the three large datasets \texttt{A, B, C},
all run on the Spark Cluster.

The size of a file in these datasets, in most cases, ranges from 200MB
to 1GB. In the implementations of all methods, each file's data form
a processing unit (an RDD in Spark), resulting in coarse-grained
parallelism. In both the baseline and \texttt{CD} methods, each
machine in the cluster automatically grabs the to-be-processed RDDs
one after another, processes them, and finally merges the results. The two versions differ in whether an RDD is formed on the
uncompressed or compressed data, and how an RDD is processed. Because
the total size of the uncompressed datasets \texttt{B} and \texttt{C}
exceeds the aggregate memory of the cluster, a newly-loaded RDD
reuses the memory of an already-processed RDD.



\texttt{Word count} and \texttt{sort} use the preorder
  traversal, \texttt{inverted index} and \texttt{term vector} use the
  postorder traversal, and \texttt{sequence count} and \texttt{ranked
    inverted index} use the depth-first traversal and the two-level
  table design of Guideline IV in
  Section~\ref{sec:treatment2orderSen}. Because the three datasets all
  consists of very large files, the data-sensitivity of order
  selection does not affect our methods of processing.\footnote{
  Section~\ref{sec:detailBen} shows the sensitivity on the other two datasets, D and E.}  
  All the programs
  use the coarse-grained parallelization. For the coarsening
  optimization, \texttt{word count}, \texttt{sort}, \texttt{inverted
    index}, and \texttt{term vector} use \emph{edge merging}, because
  they do not need to keep the order of words. \texttt{Sequence count}
  and \texttt{ranked inverted index} use \emph{node coarsening}, because
  node coarsening reduces the number of substrings spanning across
  nodes, thereby increasing the reuse of local data. We empirically set 100
  as the node coarsening threshold such that each node
  contains at least 100 items (subrules and words) after coarsening.

The average speedups with our \texttt{CD} method are \ZavgSpeedupAadd{}X,
\ZavgSpeedupBadd{}X, and \ZavgSpeedupCadd{}X on the three datasets.
\texttt{Inverted index} and \texttt{term vector} show the
largest speedups. These two programs are both unit sensitive,
producing analytics results for each file. \texttt{CD}
creates an RDD partition (the main data structure used in Spark) for
the compressed results of each file, but the baseline method cannot
because some of the original files exceed the size limit of an RDD
partition in Spark---further partitioning of the files into
segments and merging of the results incur some large overhead. Programs
\texttt{word count} and \texttt{sort} are neither unit sensitive nor order
sensitive. \texttt{Sort} has some extra code irrelevant to the
\texttt{CD} optimizations, and hence shows a smaller overall
speedup. Programs \texttt{sequence count} and \texttt{ranked inverted
  index} are both about word sequences in each file; the amount of
redundant computations to save is the smallest among all the
programs.

    \cred{
The \texttt{gzip} method incurs only 1-14\% \emph{slowdown} due to the extra decompression time,
and TADOC with double compression can still achieve significant performance benefits compared to the original version without compression,
    which proves the effectiveness of our double compression techniques.
    }

\begin{figure}[!h]
  \vspace{-0.15in}
  \center
  \includegraphics[width=1\linewidth]{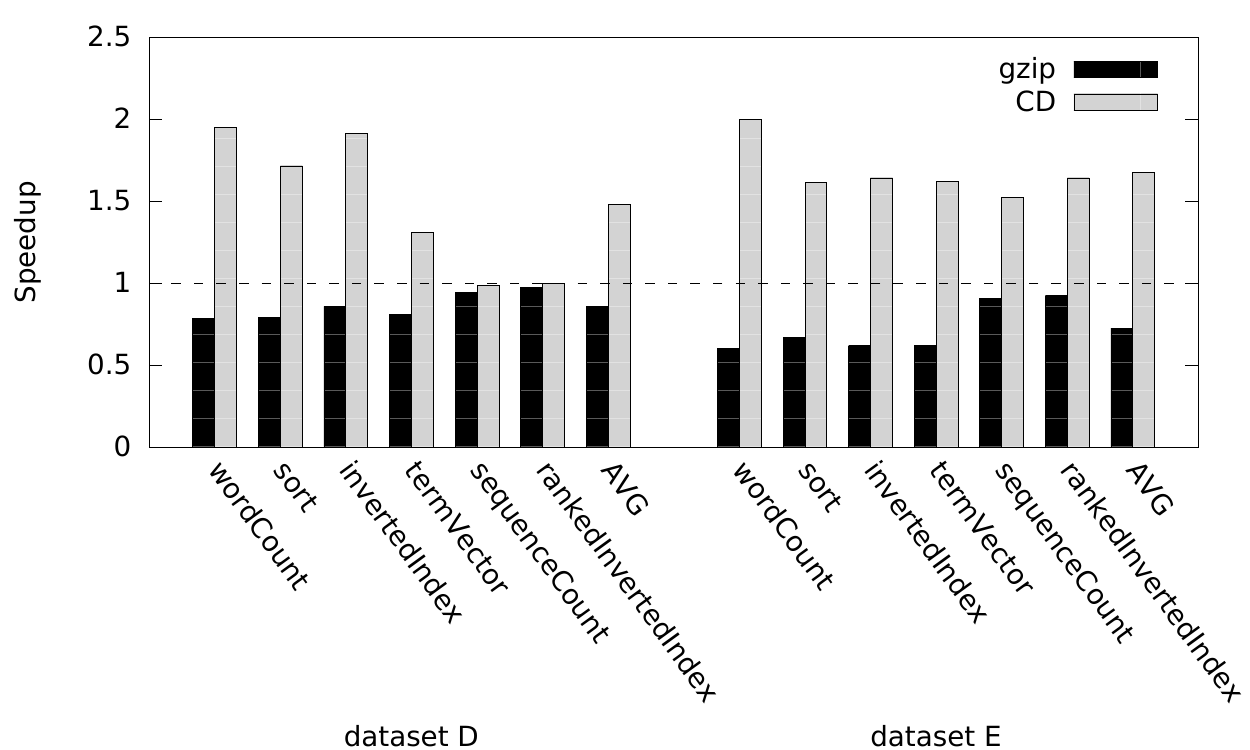}
\vspace{-0.15in}
  \caption{Performance of different methods normalized to the baseline method on the Single Node machine.}
  \label{fig:singleNode}
\vspace{-0.20in}
\end{figure}

Figure~\ref{fig:singleNode} reports the overall speedups
on the two smaller datasets on the single-node server. \texttt{CD} provides significant speedups on them as well, while the \texttt{gzip} method causes even more slowdown.
The reason is that the computation time 
on the small datasets is little and hence the decompression overhead has a more dominant effect on the overall time.
We discuss the time breakdowns in more detail next.

\begin{figure}[!h]
\vspace{-0.20in}
  \center
  \includegraphics[width=1\linewidth]{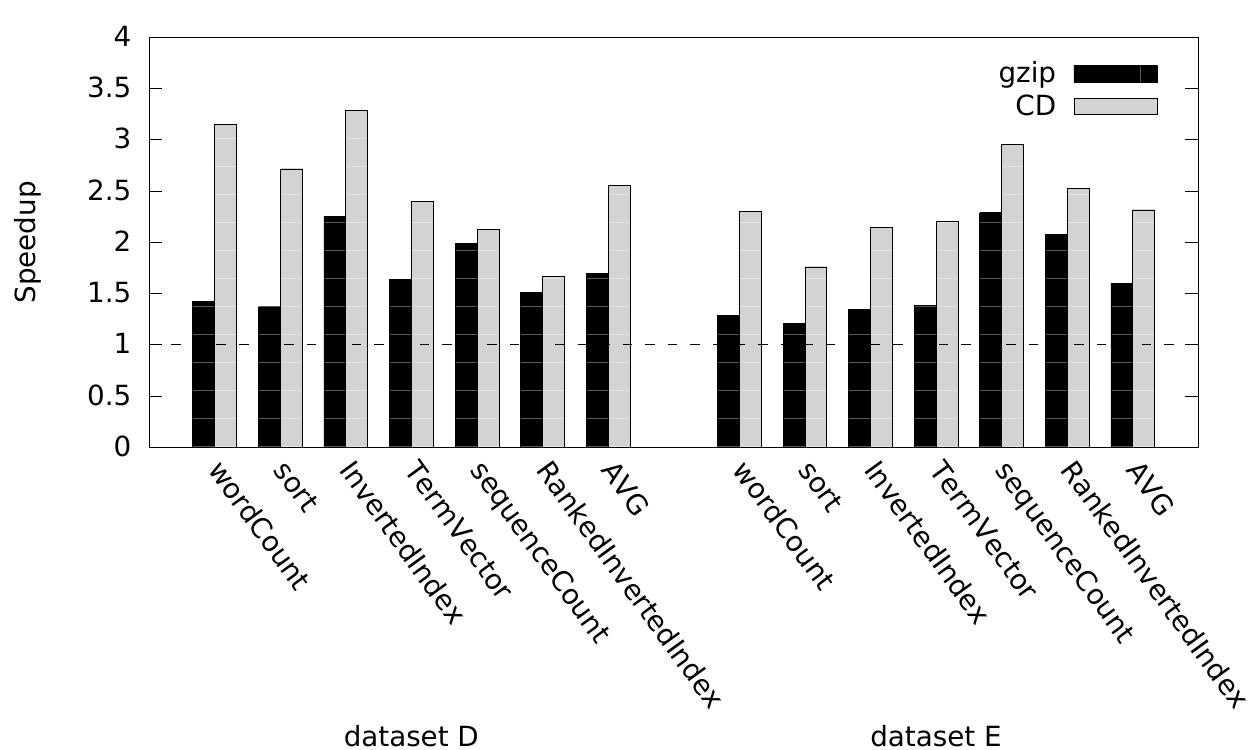}
  \vspace{-0.15in}
  \caption{\cred{
    Parallel performance of different methods normalized to the baseline method on the Single Node machine.}
    }
  \label{fig:parallelVersion}
\vspace{-0.25in}
\end{figure}

\cred{
\texttt{CD} on the single node server can also benefit from parallelism.
With coarse-grained parallelism, the computation time can be further reduced.
We implement a parallel \texttt{CD} version,
and its performance results are shown in Figure~\ref{fig:parallelVersion}.
For comparison, we also integrate the technique of coarse-grained parallelism into \texttt{gzip}.
With coarse-grained parallelism,
both \texttt{gzip} and \texttt{CD} gain performance benefits, \ 
and the average performance speedup of \texttt{CD} reaches \parallelAVG{}.
}

\vspace{-0.20in}
\subsubsection{Time Breakdowns}
\label{sec:timebd}
\vspace{-0.10in}

The right eight columns in Table~\ref{table:sequentialPerformance} report the time breakdowns on datasets D and C, 
the smallest and the largest ones.
Execution on D happens on the Single-Node server and that on C on the Spark Cluster.
The time breakdown shows that \texttt{CD} experiences
a much shorter I/O time than \texttt{gzip} does. This is because
\texttt{CD} needs to load only the compressed data into memory while
\texttt{gzip} needs to load the decompressed data.
  Although the data loading time of \textbf{CD} is shorter than that of the \textbf{gzip} method,
  the data loading time only occupies a small proportion of the whole execution time.
  The major benefits of TADOC come from the effective data reuse.

\begin{table*}[!h]
\centering
\caption{Time breakdown (seconds) and memory savings.}
\vspace{-0.10in}
\label{table:sequentialPerformance}
\resizebox{\linewidth}{!}{
\begin{tabular}{llrcrrrrrrrr}
\hline
                           &                       & \multicolumn{2}{c}{Memory}                       & \multicolumn{2}{c}{I/O Time}                          & \multicolumn{2}{c}{Init Time}                          & \multicolumn{2}{c}{Compute Time}                       & \multicolumn{2}{c}{Total Time}                    \\ \hline
                           & Benchmark             & gzip (MB) & \multicolumn{1}{c|}{CD savings (\%)} & \multicolumn{1}{c}{gzip} & \multicolumn{1}{c|}{CD}    & \multicolumn{1}{c}{gzip} & \multicolumn{1}{c|}{CD}     & \multicolumn{1}{c}{gzip} & \multicolumn{1}{c|}{CD}     & \multicolumn{1}{c}{gzip} & \multicolumn{1}{c}{CD} \\ \cline{2-12} 
 & word count            & 1157.0       & \multicolumn{1}{c|}{88.8}          & 4.0                      & \multicolumn{1}{r|}{2.6}   & 14.1                     & \multicolumn{1}{r|}{4.5}    & 0.4                      & \multicolumn{1}{r|}{0.3}    & 18.5                     & 7.4                    \\
 \multirow{1}{*}{dataset D}                          & sort                  & 1143.0       & \multicolumn{1}{c|}{88.7}          & 4.0                      & \multicolumn{1}{r|}{2.6}   & 15.0                     & \multicolumn{1}{r|}{6.1}    & 0.4                      & \multicolumn{1}{r|}{0.3}    & 19.4                     & 8.9                    \\
 data size: 0.9 GB                          & inverted index        & 1264.7       & \multicolumn{1}{c|}{79.5}          & 4.0                      & \multicolumn{1}{r|}{2.6}   & 13.4                     & \multicolumn{1}{r|}{4.0}    & 11.1                     & \multicolumn{1}{r|}{6.2}    & 28.5                     & 12.8                   \\
 CD size: 132 MB                          & term vector           & 1272.1       & \multicolumn{1}{c|}{71.0}          & 4.0                      & \multicolumn{1}{r|}{2.6}   & 13.3                     & \multicolumn{1}{r|}{7.4}    & 4.1                      & \multicolumn{1}{r|}{3.3}    & 21.4                     & 13.2                   \\
 storage saving: 84.7\%                          & sequence count        & 1734.3       & \multicolumn{1}{c|}{47.3}          & 4.0                      & \multicolumn{1}{r|}{2.6}   & 13.8                     & \multicolumn{1}{r|}{4.1}    & 50.4                     & \multicolumn{1}{r|}{58.3}   & 68.1                     & 65.0                   \\
                           & ranked inverted index & 1734.3       & \multicolumn{1}{c|}{47.3}          & 4.0                      & \multicolumn{1}{r|}{2.6}   & 13.9                     & \multicolumn{1}{r|}{4.4}    & 138.7                    & \multicolumn{1}{r|}{141.5}  & 156.6                    & 148.4                  \\ \hline
 & word count            & 177920.0     & \multicolumn{1}{c|}{89.5}          & 571.5                    & \multicolumn{1}{r|}{131.5} & 3120.0                   & \multicolumn{1}{r|}{840.0}  & 900.0                    & \multicolumn{1}{r|}{780.0}  & 4591.5                   & 1751.5                 \\
 \multirow{1}{*}{dataset C}                          & sort                  & 177920.0     & \multicolumn{1}{c|}{89.5}          & 511.5                    & \multicolumn{1}{r|}{131.5} & 2940.0                   & \multicolumn{1}{r|}{780.0}  & 1500.0                   & \multicolumn{1}{r|}{1200.0} & 4951.5                   & 2111.5                 \\
 data size: 144.4 GB                         & inverted index        & 180638.0     & \multicolumn{1}{c|}{88.1}          & 596.1                    & \multicolumn{1}{r|}{120.0} & 4140.0                   & \multicolumn{1}{r|}{600.0}  & 1380.0                   & \multicolumn{1}{r|}{480.0}  & 6116.1                   & 1200.0                 \\
  CD size: 11 GB                         & term vector           & 184138.0     & \multicolumn{1}{c|}{86.5}          & 571.5                    & \multicolumn{1}{r|}{131.5} & 3540.0                   & \multicolumn{1}{r|}{660.0}  & 1560.0                   & \multicolumn{1}{r|}{780.0}  & 5671.5                   & 1571.5                 \\
  storage saving: 92.4\%                          & sequence count        & 205117.8     & \multicolumn{1}{c|}{77.6}          & 672.9                    & \multicolumn{1}{r|}{320.0} & 5820.0                   & \multicolumn{1}{r|}{3780.0} & 1380.0                   & \multicolumn{1}{r|}{1500.0} & 7872.9                   & 5600.0                 \\
                           & ranked inverted index & 205117.8     & \multicolumn{1}{c|}{77.6}          & 672.9                    & \multicolumn{1}{r|}{260.0} & 7020.0                   & \multicolumn{1}{r|}{5280.0} & 3600.0                   & \multicolumn{1}{r|}{3480.0} & 11292.9                  & 9020.0                 \\ \hline
\end{tabular}
  }
\vspace{-0.15in}
\end{table*}

Even if I/O time is not counted,
\texttt{CD} still outperforms \texttt{gzip} substantially.
This is reflected in \texttt{CD}'s shorter times in all other parts of the time
breakdowns. For instance, \texttt{CD}'s initialization step takes about 1/3 to
1/2 of that of \texttt{gzip}. This is because 
\texttt{gzip} requires significant time to produce the completely decompressed data.

In most cases, the actual data processing part of \texttt{CD} (i.e., the
``compute time'' column in Table~\ref{table:sequentialPerformance}) is
also much shorter than that of \texttt{gzip}, thanks to
\texttt{CD}'s avoidance of the repeated processing of content that appears
\emph{multiple} times in the input datasets.\footnote{The processing time in 
\texttt{gzip} is the same as in the baseline method since
they both process the decompressed data.
}
The exceptions are
\texttt{sequence count} and \texttt{ranked inverted index} on dataset
\texttt{D}. These two programs are both unit and order sensitive.
Dataset \texttt{D}, which consists of
many small files, does \emph{not} have many repeated word sequences,
so obtaining performance improvement on it is even harder.
However,
even for these two extreme cases, the overall time of \texttt{CD} is
still shorter than that of \texttt{gzip} because of \texttt{CD}'s substantial
savings in the I/O and initialization steps.
We conclude that our \texttt{CD} method greatly reduces execution time on many workloads and datasets.



\vspace{-0.20in}
\subsection{Space Savings}
\label{sec:space}
\vspace{-0.10in}

Table~\ref{table:sec6compression} reports the compression ratio, 
which is
defined as {\it size(original)/size(compressed)}. In all 
methods that use compression,
the datasets are already dictionary-encoded. 
Compression methods apply to both the datasets and the dictionary. The
CD- row shows the compression ratios from Sequitur alone.
Sequitur's compression ratio is
2.3--3.8, considerably smaller than the ratios from Gzip. However, with
the double compression technique, \texttt{CD}'s compression ratio is
boosted to 6.5--14.1,
which is greater than the Gzip ratios.
Gzip results \emph{cannot} be used for direct data processing, but Sequitur
results can, which enables \texttt{CD} to bring significant time
savings as well, as reported in Section~\ref{sec:speedups}.

\begin{table}[!h]
\centering
\caption{Compression ratios.}
\vspace{-0.10in}
\begin{tabular}{l|lllll|l}\hline
            & \multicolumn{5}{c|}{Dataset} &  \\
Version      & A    & B    & C    & D    & E    & AVG  \\ \hline
default      & 1.0  & 1.0  & 1.0  & 1.0  & 1.0  & 1.0 \\
gzip         & 9.3  & 8.9  & 8.5  & 5.9  & 8.9  & 8.3 \\
CD           & 14.1 & 13.3 & 13.1 & 6.5  & 11.9 & 11.8 \\
CD-          & 3.1  & 3.2  & 3.8  & 2.3  & 2.8  & 3.0 \\\hline
\end{tabular}\\
\scriptsize{CD-: Sequitur without double compression.}
\label{table:sec6compression}
\end{table}

The ``Memory'' columns in Table~\ref{table:sequentialPerformance} report the memory savings by ``\texttt{CD}'' compared to the memory usage by the
\texttt{gzip} method. 
Because \texttt{CD} loads and processes much less data, it reduces memory usage
by \memSavingAVG{}. This benefit is valuable considering the large
pressure modern analytics pose to the memory space of modern
machines. The smaller memory footprint also helps \texttt{CD} to reduce memory access
times.

\vspace{-0.20in}
\subsection{\cred{Evaluation of Advanced Document Analytics}}
\vspace{-0.10in}

\subsubsection{\cred{Time Benefits}}
\begin{figure}[!h]
\vspace{-0.15in}
  \center
  \includegraphics[width=1\linewidth]{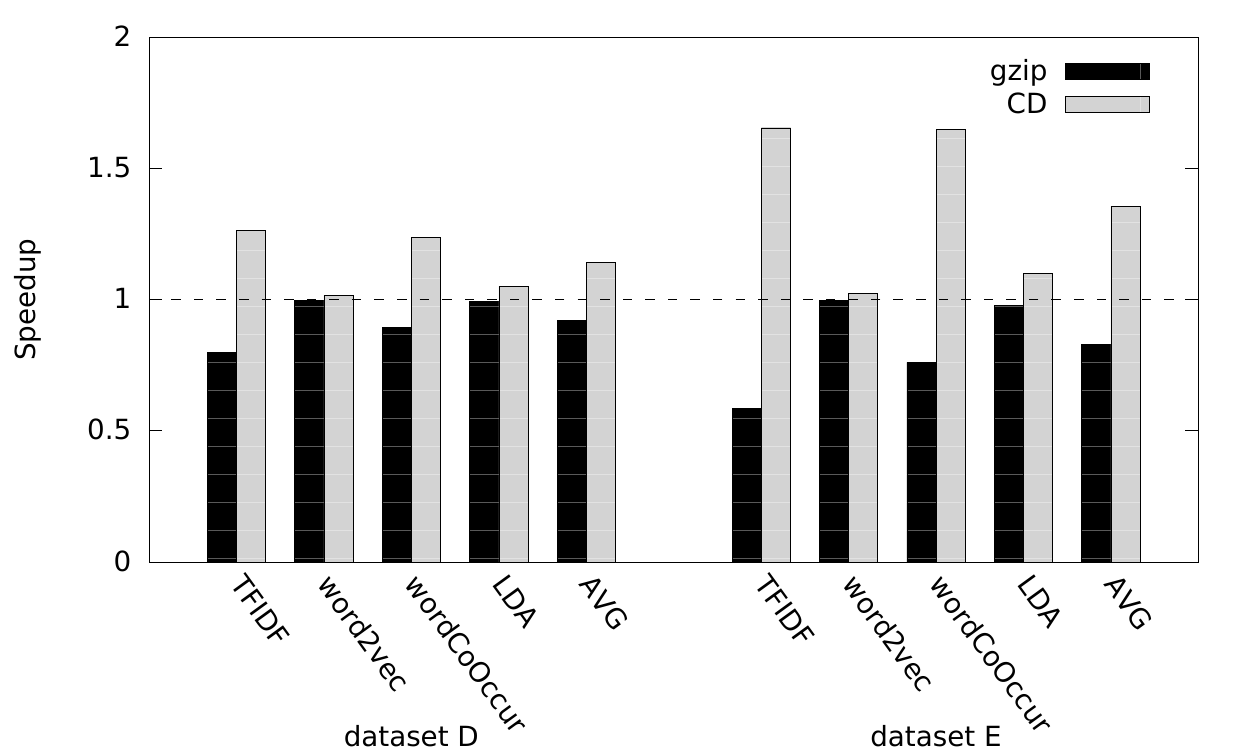}
\vspace{-0.15in}
  \caption{
\cred{
    Performance of different methods normalized to the baseline of the advanced document analytics applications.}
}
  \label{fig:advancedSpeedup}
\vspace{-0.35in}
\end{figure}

\vspace{-0.10in}
\cb{
In this section, we measure the performance of TADOC in the four applications described in Section~\ref{sec:advanced}.
We show the execution time benefits in Figure~\ref{fig:advancedSpeedup}.
As introduced in Section~\ref{subsec:evaluationEva},
the baseline is the execution time of the original implementation,
which directly processes the original non-compressed dataset.
\texttt{gzip} uses Gzip to compress the original data, and during processing time, it needs to decompress the compressed data before data processing.
\texttt{CD} is the version using TADOC, which also includes a gunzip stage.
We use the speedup over the baseline as the metric to quantify time benefits.
The \texttt{gzip} version suffers from data decompression overheads,
so we focus on the analysis of the \texttt{CD} version of TADOC.
}

\cred{
In Figure~\ref{fig:advancedSpeedup},
\texttt{CD} yields 1.2X speedup on average over the baseline.
Among the four applications, TFIDF experiences a performance benefit greater than 30\%.
The reason for this result is that these applications can be divided into two stages, namely, data preprocessing and processing,
and TADOC is mainly used in the data preprocessing stage;
the preprocessing stage may account for different proportions of the total execution time.
The time breakdown is shown in Table~\ref{tbl:advancedBreakdown},
where the last column represents the proportion of preprocessing time.
Data preprocessing accounts for 44.6\% (\texttt{D}) and 64.7\% (\texttt{E}) of execution time  for TFIDF.
For word co-occurrence, data preprocessing accounts for 35.4\% (\texttt{D}) and 21.3\% (\texttt{E}) of execution time.
The preprocessing proportion of these two applications is high compared to that of the other applications,
and additionally,
the decompression process of \texttt{CD} is shorter than that of \texttt{gzip}.
Therefore, TFIDF and word co-occurrence have relatively high performance speedups.
}

\begin{table}[!h]
\centering
  \caption{\cred{Time breakdown for advanced applications.}}
\vspace{-0.10in}
\resizebox{\linewidth}{!}{%
\begin{tabular}{ccccc}
\hline
  & Benchmark        & Preprocessing  & Total & Occupancy \\ 
  &                  & (s) & Time (s) & (\%) \\ \hline
  D & TFIDF            & 5.6  & 12.6   & 44.6 \\
  & wordCoOccur & 9.6  & 27.2   & 35.4 \\
  & LDA              & 20.3 & 448.9  & 4.5  \\
  & word2vec         & 7.0  & 2082.6 & 0.3  \\ \hline
E & TFIDF            & 7.4  & 11.5   & 64.7 \\
  & wordCoOccur & 5.5  & 25.9   & 21.3 \\
  & LDA              & 14.9 & 502.9  & 3.0  \\
  & word2vec         & 7.0  & 2864.0 & 0.2 \\ \hline
\end{tabular}%
}
  \label{tbl:advancedBreakdown}
\vspace{-0.20in}
\end{table}

\cred{
In general,
when the data preprocessing and decompression stages account for a high proportion of the total execution time,
TADOC has significant performance benefits;
otherwise, the execution time of TADOC is lower than but closer to the original execution time.
}

\subsubsection{
\cred{
Storage and Memory Benefits
}}
\vspace{-0.15in}
\cb{
We evaluate the storage and memory benefits of TADOC on advanced analytics workloads.
}
\cred{
Storage savings are the same as those in the results in Section~\ref{sec:space}.
Compared to the original datasets,
\texttt{CD} reaches 6.5 and 11.9 compression ratios for datasets \texttt{D} and \texttt{E},
respectively,
which implies that TADOC brings more than 90\% storage reduction.
}

\cred{
Memory savings are shown in Table~\ref{tbl:advancedMem}
and range from 0.6\% to 76.2\%.
They vary considerably
because different applications use different auxiliary data structures.
For example,
for TFIDF,
the algorithm is simple, and
we only need to develop data structures for storing the TF and IDF.
In contrast,
for LDA,
although the ``bag-of-words'' paradigm used in preprocess relies on only word frequencies,
}
the training stage involves \cb{many intermediate data} structures in model construction,
which decreases TADOC's memory benefits.

\begin{table}[!h]
  \vspace{-0.2in}
  \caption{\cred{Memory savings for advanced applications.}}
  \vspace{-0.1in}
\centering
\resizebox{\linewidth}{!}{%
\begin{tabular}{cccc}
\hline
  & Benchmark   & Original (MB) & Memory Savings (\%) \\ \hline

  D & TFIDF       & 1617.0 & 60.5 \\
  & wordCoOccur & 1679.0 & 17.7 \\
  & LDA         & 2552.4 & 0.6 \\
  & word2vec    & 772.3  & 17.1 \\ \hline
E & TFIDF       & 1459.4 & 76.2 \\
  & wordCoOccur & 1881.0 & 15.2 \\
  & LDA         & 3883.6 & 3.4  \\
  & word2vec    & 911.0  & 18.4 \\ \hline

\end{tabular}%


}
  \label{tbl:advancedMem}
  \vspace{-0.2in}
\end{table}


\vspace{-0.20in}
\subsection{When Inverted Index is Used}
\label{sec:cdWithInvertedIdx}
\vspace{-0.10in}

In some cases, practitioners store an inverted index~\cite{zhang2001supporting,whang2002inverted,mitsui1993information} with the original dataset.
Doing so helps accelerate some analytics tasks.
This approach can be combined with TADOC by attaching an inverted index of the original documents with the Sequitur compression result.
We call these two schemes Original+index and CD+index.
For tasks where inverted index can be used (e.g., the first four benchmarks),
some intermediate results can be obtained directly from inverted index to save time.
For the other tasks (e.g., \texttt{sequence count}, \texttt{ranked inverted index}), Original+index has to fall back to the original text for analysis,
and CD+index provides 1.2X-1.6X speedup due to its direct processing on the Sequitur DAG.
Besides its performance benefits, CD+index saves about 90\% space over the Original+index as Table~\ref{tbl:spaceCost} reports.

\begin{table}[!h]
\vspace{-0.20in}
\centering
  \caption{Space usage of the original datasets and CD with inverted-index.}
\vspace{-0.10in}
\label{tbl:spaceCost}
\resizebox{\linewidth}{!}{
\begin{tabular}{ccrrrrr}
\hline
Usage        & Dataset        & \multicolumn{1}{c}{A} & \multicolumn{1}{c}{B} & \multicolumn{1}{c}{C} & \multicolumn{1}{c}{D} & \multicolumn{1}{c}{E} \\ \hline
Memory  & Original+Index & 32,455                 & 92,234                 & 184,469                & 1,387                  & 1,406                  \\
(MB)    & CD+Index       & 3,693                  & 10,405                 & 20,806                 & 413                   & 197                   \\ \hline
Storage & Original+Index & 37,990                 & 78,438                 & 154,214                & 1,115                  & 1,559                  \\
(MB)    & CD+Index       & 2,873                  & 6,066                  & 11,965                 & 211                   & 140                   \\ \hline
\end{tabular}
\vspace{-0.20in}
}
\end{table}

\vspace{-0.40in}
\subsection{More Detailed Benefits}
\label{sec:detailBen}
\vspace{-0.10in}

In this part, we briefly report the benefits coming from each of the major
guidelines we described in Section~\ref{sec:solutions}.

\begin{figure}[!h]
\vspace{-0.2in}
 \centering                            
   \includegraphics[width=1\linewidth]{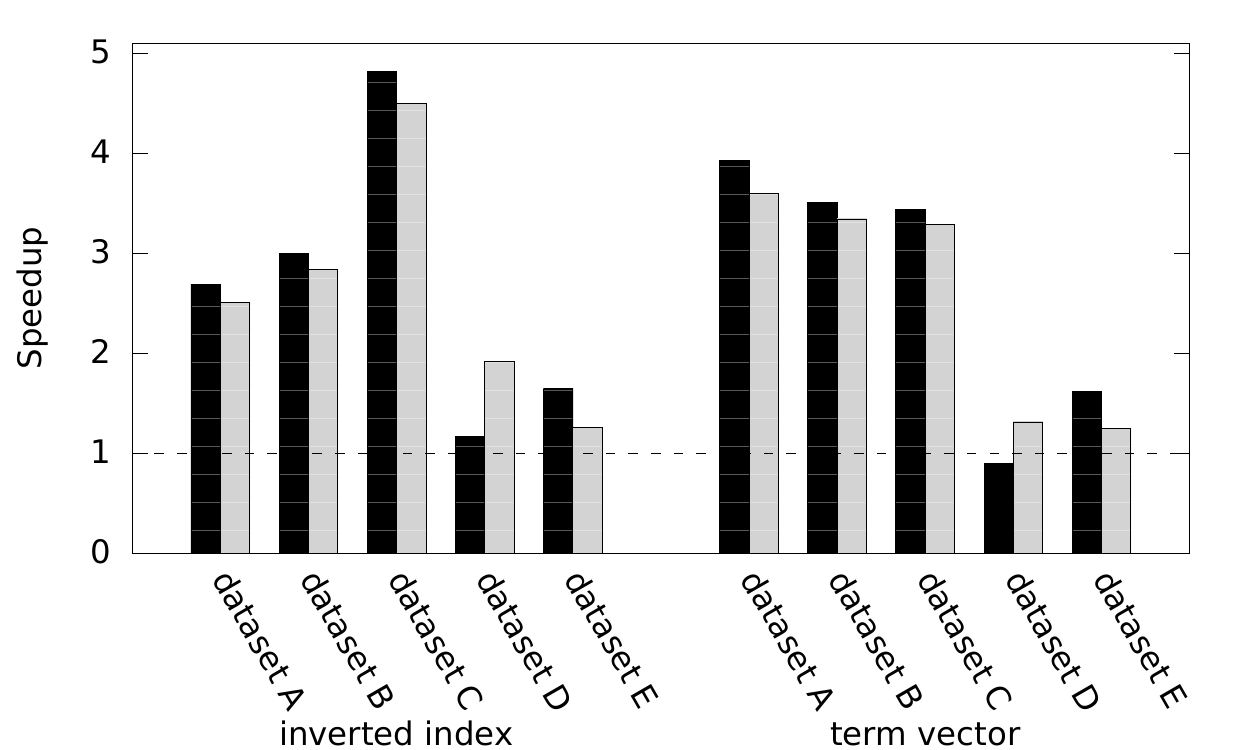}
\vspace{-0.15in}
   \caption{Performance of preorder and postorder for \texttt{inverted index} and \texttt{term vector}.}
 \label{fig:preposttermvector}
\vspace{-0.20in}
 \end{figure}

The benefits of {\em adaptive traversal order} (Guideline I and II) are most prominent on benchmarks
\texttt{inverted index} and \texttt{term vector}. 
Using adaptive traversal order,
the \texttt{CD} method selects postorder traversal when processing dataset \texttt{D} and
preorder on datasets \texttt{A, B, C, E}. 
We show the performance of both preorder and postorder traversals for \texttt{inverted index} and \texttt{term
vector} in Figure~\ref{fig:preposttermvector}. Using decision trees, 
\texttt{CD} successfully selects the better traversal order for each of the
  datasets. For instance, on \texttt{inverted index},
\texttt{CD} picks postorder on dataset \texttt{D}, which
outperforms preorder by 1.6X, and it picks preorder on dataset \texttt{E}, which outperforms postorder by 1.3X.

{\em Double compression} (Guideline VI) provides substantial space benefits as we
have discussed in Section~\ref{sec:forCompression}. 
However, since double compression needs to recover the
Sequitur results from the compressed data before processing, it incurs some
overhead. Our experiments show that this overhead is 
outweighed by the overall time benefits of \texttt{CD}.

We tried to implement a fine-grained parallel version of \texttt{CD} for
benchmark \texttt{word count}. It breaks the CFG into a number of partitions and
uses multiple threads to process each partition in parallel. 
Even though this version
took us several times the effort we spent on the {\em
  coarse-grained parallel} version (Guideline III), its performance was substantially
worse (e.g., 50\% slower on dataset \texttt{D}).

{\em Double-layered bitmap} (Guideline V) helps \emph{preorder} processing on datasets that contain many (>800) files of medium size (>2860 words per Figure~\ref{fig:decisionTree}.) Among the 60 datasets involved in the decision tree experiments in Section~\ref{sec:order}, 10 of them works best with double-layered bitmap based preorder. 
  They get 2\%-10\% performance benefits compared to single-layered bitmap based preorder traversal. 
  Besides double-layered bitmap, we experiment with other alternative data structures, including \emph{red-black tree}~\cite{cormen2009introduction}, \emph{hash set}~\cite{cormen2009introduction}, and \emph{B-tree}~\cite{btreeImplementation}. Table~\ref{comparison2lev} reports the performance of preorder \texttt{inverted index} when these data structures are used in place of double-layered bitmap in each of the DAG node.
The experiment uses dataset D and the Single-Node server in Table~\ref{configurations}. 
Double-layered bitmap is fast to construct as it uses mainly bit operations.
The query time for double-layered bitmap has a complexity of $O(1)$.
Some of the alternative data structures (e.g., B-tree) yield a shorter processing time,
but suffer a longer construction process (i.e., initialization in Table~\ref{comparison2lev}).



\begin{table}[!h]
\vspace{-0.2in}
\centering
  \caption{Performance and time breakdown of different data structure achieves for inverted-index.}
\vspace{-0.10in}
\label{comparison2lev}
\resizebox{\linewidth}{!}{
\begin{tabular}{cccc}
\hline
Data Structure & Initialization (s) & Computation (s) & Total (s) \\ \hline
  2LevBitMap   & 14.96             & 3.08           & \bf{18.04}    \\
  redBlackTree & 39.33             & 3.56           & \bf{42.89}    \\
  hash set      & 25.34             & 4.32           & \bf{29.67}    \\
  B-tree        & 18.87             & 2.29           & \bf{21.16}    \\ \hline
\end{tabular}
}
\vspace{-0.25in}
\end{table}

Finally, \emph{coarsening} (Guideline VI) shows clear benefits for \texttt{CD} on benchmarks
\texttt{ranked inverted index} and \texttt{sequence count}. For
instance, compared to no coarsening, it enables the CD-based \texttt{ranked
  inverted index} program to achieve 5\% extra performance improvement on dataset
\texttt{E}. The benefits of Guideline IV
has been reported in Section~\ref{sec:treatment2orderSen} and are hence omitted here.

\vspace{-0.2in}
\subsection{Compression Time and Applicability}
\label{sec:compressTimeApp}
\vspace{-0.10in}

   \cred{
The time taken to compress the datasets using sequential Sequitur ranges from 10 minutes to over 20 hours.
Using a parallel or distributed Sequitur with accelerators (e.g., as in~\cite{DBLP:conf/asplos/BoroumandGKASTK18,DBLP:conf/micro/PekhimenkoSKXMGKM13})  can potentially shorten the compression time substantially.
Note that this article focuses on how to use the compressed data to support various analytics tasks such as  word count;
we do not further discuss the compression process.
}

\cred{
  In general, our technique is designed for document analytics that can be expressed as a DAG traversal-based problem on datasets that do \emph{not} change frequently. 
 Moreover,
our discussion has focused on applications that normally require scanning the entire dataset,
as illustrated by the analytics problems used in our evaluation.
  It is not designed for regular expression queries or scenarios where data frequently changes. We note that the proposed technique can also benefit advanced document analytics. The initial part of many advanced document analytics is to load documents and derive some representations (e.g., natural language understanding) to characterize the documents such that later processing can efficiently work on these representations. One example is email classification, where TADOC can help to accelerate the construction of the feature vectors (e.g., word frequency vectors) required for classification.
    For the applications that cannot be expressed as a DAG traversal-based problem, 
    TADOC can be used as a storage technique.
For example, to compute Levenshtein distance between a pair of words~\cite{levenshtein1966binary},
    the required words can be extracted from the compressed data~\cite{9101809},
    and then, the application of Levenshtein edit distance can be performed on the extracted words.
}

\vspace{-0.20in}
\section{Related Work}
\label{sec:related}
\vspace{-0.10in}

To our knowledge, this is the first work to enable \emph{efficient} direct document analytics on compressed data.
The work closest to \texttt{CompressDirect} is
Succinct~\cite{agarwal2015succinct,khandelwal2016blowfish}, which enables efficient queries
on compressed data in a database. 
These two techniques are complementary to each other. Succinct is based on index and suffix
array~\cite{navarro2016compact}, an approach employed in other 
works as well~\cite{agarwal2015succinct,burrows1994block,ferragina2005indexing,grossi2004indexing,ferragina2009compressed}.  \texttt{CompressDirect} and these previous
studies differ in both their applicability and main techniques.
First, Succinct is mainly for the database domain while \texttt{CompressDirect} is for general document analytics. 
Succinct is designed mainly for search and random access of local queries.
Even though it could possibly be made to work on some of the general document analytics tasks, its efficiency is much less than \texttt{CompressDirect} on such tasks as those tasks are not its main targets. For instance,
\texttt{word count} on dataset \texttt{E} takes about 230 seconds with Succinct,
but only 10.3 seconds with \texttt{CompressDirect}, on the single node machine in Table~\ref{configurations}.
%
%
Second, Succinct and \texttt{CompressDirect} use different compression
methods and employ different inner storage structures. Succinct
compresses data in a flat manner, while \texttt{CompressDirect} uses Sequitur to create a
DAG-like storage structure. The DAG-like structure allows \texttt{CompressDirect} to efficiently
perform general computations for all items in the documents,
even in the presence of various challenges from files or
word sequences, as described in Section~\ref{sec:complexity}.

\cred{
  Note that Sequitur can be replaced by other context-free compression techniques.
  For example,
Re-Pair, proposed by Larsson and Moffat~\cite{larsson2000off},
is an offline dictionary-based compression algorithm.
Re-Pair can be regarded as a compromise in terms of compression time and compression ratio,
which could be a potential alternative compression algorithm.
Larson and Moffat~\cite{reparurl} offer a high-quality Re-Pair implementation.
    Ga{\'n}czorz and others~\cite{ganczorz2017improvements} further improve the Re-Pair grammar compressor by involving penalties.
}

\cred{
Traditional approaches to compression-based analytics use indexes, suffix
arrays, and suffix trees~\cite{navarro2016compact,agarwal2015succinct,burrows1994block,ferragina2005indexing,grossi2004indexing,ferragina2009compressed,farruggia2014bicriteria,ferragina2009bit,ferragina2009compressed,gog2014theory}.
Suffix trees~\cite{navarro2016compact,sadakane2007compressed}
are traditional compact data structures;
they consume less storage space while enabling analytics on compressed data.
However, research~\cite{hon2004practical,kurtz1999reducing}
shows that optimized representations consume larger memory even more than the size of the input.
Burrows-Wheeler Transform~\cite{ferragina2005indexing,burrows1994block} and suffix arrays~\cite{manber1993suffix} are milestones in the development of compact representations,
but experiments~\cite{hon2004practical} still show that they cannot solve the large memory consumption issue.
FM-indexes~\cite{wikipediaFMIndex,ferragina2000opportunistic,ferragina2001experimental,ferragina2001experimental2,ferragina2005indexing} and Compressed Suffix Array~\cite{grossi2003high,grossi2005compressed,sadakane2000compressed,sadakane2002succinct,sadakane2003new} are two efficient alternatives that further reduce the memory space consumption,
and, based on these technologies,
Agarwal and others propose Succinct~\cite{agarwal2015succinct}.
Our method, TADOC, is different from these methods; we provide a grammar-based approach, which provides new insight to 
the domain of compression-based data analytics.
}

\cred{
There are many works on grammar compression and operations on grammar-encoded strings~\cite{rytter2004grammar,charikar2005smallest,gagie2012faster,bille2015random,bille2015finger,brisaboa2019gract,ganardi2019balancing,takabatake2017space}.
Rytter~\cite{rytter2004grammar} survey the complexity issues involved in grammar compression, LZ-Encodings, and string algorithms.
Charikar and others~\cite{charikar2005smallest} study the limit of the smallest context-free grammar to generate a given string
and analyze the bound approximation ratios for well-known grammar-based compression algorithms, including Sequitur.
Gagie and others~\cite{gagie2012faster} develop a novel grammar-based self-indexing method for highly-repetitive strings such as DNA sequences.
Bille and others~\cite{bille2015random} study how to perform random access to grammar-compressed data.
    SOLCA~\cite{takabatake2017space} is a novel fully-online grammar compression algorithm that 
    targets online use cases.
}

\cred{
There is a large body of literature on inverted index compression.
Petri and Moffat~\cite{petri2018compact} explore general-purpose compression tools for compact inverted index storage.
 Moffat and Petri~\cite{moffat2018index} apply two-dimensional contexts and the byte-aligned entropy coding of asymmetric numeral systems to index compression.
Pibiri and others~\cite{pibiri2019fast} develop a fast dictionary-based compression approach, DINT, for inverted indexes.
Pibiri and Venturini~\cite{pibiri2019techniques} survey the encoding algorithms for inverted index, including single integers,
a sorted list of integers, and the inverted index.
Pibiri and others~\cite{pibiri2020compressed} propose compressed indexes for fast searching of large RDF datasets.
Oosterhuis and others~\cite{oosterhuis2018potential} apply the recursive graph bisection in document reordering,
which is an essential preprocessing phase in building indexes.
Furthermore, Mackenzie and others~\cite{mackenzie2019compressing} use machine learned models to replace common index data structures.
These works mainly consider how to compress the inverted index.
Different from inverted index compression,
TADOC focuses on how to perform analytics directly on compressed data without decompression, such as building inverted indexes directly on compressed data.
}

\cred{
Since its development~\cite{nevill1997compression,nevill1997identifying,nevill1997linear},
Sequitur has been applied to various tasks, including program and data pattern analysis~\cite{chilimbi2001efficient,chilimbi2002dynamic,larus1999whole,lau2005motivation,law2003whole,lin2005supporting,walkinshaw2010using}.
Lau and others~\cite{lau2005motivation} use Sequitur in code analysis to search
some program patterns.  Chilimbi~\cite{chilimbi2001efficient} uses
Sequitur as a representation to quantify and exploit data reference
locality for program optimization. Larus~\cite{larus1999whole}
proposes an approach called whole program paths (WPP), which
leverages Sequitur to capture and represent dynamically executed control
flow.  Law and others~\cite{law2003whole} propose a whole program
path-based dynamic impact analysis and related compression based on
Sequitur.  Chilimbi and others~\cite{chilimbi2002dynamic} use Sequitur
for fast detection of hot data streams.
Walkinshaw and others~\cite{walkinshaw2010using} apply Sequitur to the comprehension
of program traces at varying levels of abstraction. 
Lin and others~\cite{lin2005supporting} extend Sequitur as a new XML
compression scheme for supporting query processing.
We are not aware
of prior usage of Sequitur to support direct document analytics on
compressed data, as we propose.
}

\vspace{-0.20in}
\section{Conclusion}
\label{sec:conclusion}
\vspace{-0.10in}

\cred{
We propose a new method, TADOC, 
to enable high performance document analytics on
compressed data. By enabling efficient direct processing on compressed
data, our method reduces  storage space by \reduceRatioAVG{} and
 memory usage by \memSavingAVG{},
while also speeding up the analytics by \singleAVG{} on sequential systems, and \distriAVG{} on distributed clusters.
We present how the proposed method can be
materialized on Sequitur, a compression method that produces
hierarchical grammar-like representations. 
}
We discuss the major
challenges in applying the method to various document analytics tasks,
and provide a set of guidelines for developers to avoid potential
pitfalls in applying TADOC. In addition, we produce a library
named \texttt{CompressDirect} to help ease the required
development effort in using TADOC. Our results demonstrate the promise of TADOC in various environments, ranging from sequential
to parallel and distributed systems.

\vspace{-0.20in}

\begin{acknowledgements}
  This work is supported by the National Key R\&D Program of China (Grant No. 2017YFB1003103), 
  National Natural Science Foundation of China (No. 61732014 and 61802412)
  Beijing Natural Science Foundation (No. 4202031 and L192027), Tsinghua University Initiative Scientific Research Program (20191080594), and \cg{Beijing Academy of Artificial Intelligence (BAAI)}.
  Onur Mutlu is supported by ETH Z{\"u}rich, SRC, and various industrial partners of the SAFARI Research Group,
including Alibaba, Huawei, Intel, Microsoft, and VMware.
Jidong Zhai, Xipeng Shen, and Xiaoyong Du are the corresponding authors of this paper.
\end{acknowledgements}


\vspace{-0.3in}


\bibliographystyle{abbrv}

\end{document}